\documentclass[superscriptaddress,onecolumn,secnumarabic,nobibnotes,aps,prd,showpacs,nofootinbib]{revtex4}
\usepackage{graphicx}
\usepackage{epsf}
\usepackage{bm}
\usepackage{amsmath}
\usepackage{amsfonts}
\usepackage{amssymb}
\usepackage{epstopdf}
\usepackage{color}
\usepackage{subfig}

\providecommand{\U}[1]{\protect\rule{.1in}{.1in}}

\newcommand{\be}{\begin{equation}}
\newcommand{\ee}{\end{equation}}

\newcommand{\mincir}{\raise
-3.truept\hbox{\rlap{\hbox{$\sim$}}\raise4.truept\hbox{$<$}\ }}
\newcommand{\magcir}{\raise
-3.truept\hbox{\rlap{\hbox{$\sim$}}\raise4.truept\hbox{$>$}\ }}

\begin{document}

\title{Dynamics of nonlinear interacting dark energy models}
\author{Andronikos Paliathanasis}
\email{anpaliat@phys.uoa.gr}
\affiliation{Instituto de Ciencias F\'{\i}sicas y Matem\'{a}ticas, Universidad Austral de
Chile, Valdivia, Chile}
\affiliation{Institute of Systems Science, Durban University of Technology, PO Box 1334,
Durban 4000, Republic of South Africa}
\author{Supriya Pan}
\email{supriya.maths@presiuniv.ac.in}
\affiliation{Department of Mathematics, Presidency University, 86/1 College Street,
Kolkata 700073, India}
\author{Weiqiang Yang}
\email{d11102004@163.com}
\affiliation{Department of Physics, Liaoning Normal University, Dalian, 116029, P. R.
China.}

\begin{abstract}
We investigate the cosmological dynamics of interacting dark energy models
in which the interaction function is a nonlinear in terms of the energy
densities. Considering explicitly the interaction between a pressureless
dark matter and a scalar field, minimally coupled to Einstein gravity, we
explore the dynamics of the spatially flat FLRW universe for the exponential
potential of the scalar field.. We perform the stability analysis for the
three nonlinear interaction models of our consideration through the analysis
of critical points and we investigate the cosmological parameters and we
discuss the physical behaviour at the critical points. From the analysis of
the critical points we find a number of possibilities that include the
stable late time accelerated solution, $w$CDM-like solution, radiation-like
solution and moreover the unstable inflationary solution as well.
\end{abstract}

\maketitle
\date{\today }


\section{Introduction}

\label{intro}

The dark sector of our universe, according to a series of past and latest
observational evidences \cite{snia1,
snia2,Spergel:2003cb,Tegmark:2006az,Spergel:2006hy,Ade:2015xua,Amendola:2016saw,Aghanim:2018eyx,Abbott:2018wog}%
, is composed by two heavy dark fluids, namely a pressureless dark matter
and a dark energy fluid. The former fluid is responsible for the structure
formation of the universe while the latter fluid drives the present day
accelerating phase of the universe. In addition to that, the observational
evidences also estimate that nearly 96\% of the total energy density is
coming from this joint dark sector where in particular, the dark energy
contributes around 68\% of the total energy budget of the universe while the
28\% of the total energy density of the universe is from dark matter.
However, the evolution, origin and the nature of these dark fluids of the
dark energy are not clearly understood yet. Although from indirect
gravitational effects, the nature of dark matter seems to be partially
known, however, the dark energy has remained to be extremely mysterious. As
a consequence, a number of cosmological models have been introduced and
investigated in the last couple of years. The simplest cosmological
consideration is the non-interacting models of the universe where dark
matter and dark energy are conserved separately, leading to two independent
evolutions of these dark fluids. While on the other hand, a more generalized
version of the cosmological models is available in which dark matter and
dark energy are allowed to interact with other. In the present work we shall
consider the interacting cosmological models.

The interaction between dark matter and dark energy is a potential mechanism
to explain the cosmic coincidence problem \cite%
{Amendola-ide1,Amendola-ide2,Pavon:2005yx,delCampo:2008sr,delCampo:2008jx},
although its origin was motivated to explain the discrepancy in the
cosmological constant. An earlier investigation by Wetterich \cite%
{Wetterich-ide1} shows that an interaction between a scalar field and
gravity could lead to an effective cosmological constant which is dynamical
in nature and asymptotically approaches toward a tiny value and thus the
mismatched value in the cosmological constant gains a plausible explanation.
Thus, the interaction in the dark sector started its beginning following
these two motivations.

The dark sector interaction is a phenomenological consideration because
there is no such fundamental principle that could derive it, however, from
the theoretical ground, precisely from the particle physics theory, any two
matter fields (here dark matter and dark energy fields) can interact with
each other. Interestingly, this specific phenomenological theory has gained
a massive interest in the cosmological community for several potential
outcomes. It has been found that the allowance of an interaction can take
the dark energy equation of state from quintessence to phantom regime, that
means an effective quintom type of nature is imposed in the dark energy
state parameter. The phantom crossing available in scalar field models with
negative kinetic correction (known as phantom scalar field models) lead to
instabilities at the classical and quantum levels. Secondly, the interaction
has been found to be very efficient to address the mismatched value of the
Hubble constant $H_{0}$ from the global ($\Lambda $CDM based Planck) and
local measurements. Some other investigators have also found that the
interaction might be able to solve the tension on $\sigma _{8}$.

On the other hand, from the recent observational evidences, it has been
already pointed out that irrespective of the dark energy equation of state ($%
w_{x}=-1$, $w_{x}$ is constant but $\neq -1$, or $w_{x}$ is dynamical), the
interaction in the dark sector is allowed, although the strength of the
interaction is mild, but it is not ruled out though \cite%
{Yang:2014gza,Salvatelli:2014zta,Yang:2014hea,
Nunes:2016dlj,Kumar:2016zpg,Yang:2016evp,Kumar:2017dnp,DiValentino:2017iww,Yang:2017yme}%
. Thus, based on the above observational predictions, one can assume that
the interaction in the dark sector might be a potential theory for further
investigations. For a general overview of different interaction models and
their cosmological consequences, we refer to a number of past \cite%
{Amendola-ide1,Amendola-ide2,Pavon:2005yx,delCampo:2008sr,delCampo:2008jx,Billyard:2000bh,Barrow:2006hia,Amendola:2006dg, CalderaCabral:2008bx, Quartin:2008px, Valiviita:2009nu, Clemson:2011an, Thorsrud:2012mu}
and recent works \cite{Yang:2014gza,Salvatelli:2014zta,Yang:2014hea,
Nunes:2016dlj,Kumar:2016zpg,Yang:2016evp,Kumar:2017dnp,DiValentino:2017iww,Yang:2017yme,Pan:2012ki,Pan:2016ngu,Sharov:2017iue, Pan:2017ent,Yang:2018euj, Yang:2018pej,Yang:2018ubt,int1,int2,int3,int4,int5,int6,int7}
containing some interesting observations.

Now concerning the choices of the interaction functions, as there is no such
governing rule, thus, in principle a number of linear and nonlinear
functions can be chosen. However, the models with nonlinear interactions are
rare in the literature \cite{Chimento:2009hj, Arevalo:2011hh, Yang:2017zjs},
since the dynamics of the interacting fluids becomes much complicated
compared to the dynamics for linear interaction. Nevertheless, it is always
fascinating to explore the dynamics in the context of nonlinear interaction
functions in order to see if we can extract more information out of that.
Thus, being motivated, in the present work we consider an interacting
scenario between a scalar field and the pressureless dark matter where the
interaction functions are nonlinear in nature. We have performed the
stability analysis of each interaction model in order to investigate their
cosmological viabilities. The work has been organized as follows.

In Section \ref{field-equations}, we present the gravitational equations for
an interacting universe. In Section \ref{formation}, we describe the
formation of the dynamical system for the interacting scenarios. In section %
\ref{sec-critical} we study the critical points for all the interacting
scenarios and we present the stability analysis. Finally, our discussion and
conclusions are given in\ Section \ref{sec-conclu}.

\section{Field equations}

\label{field-equations}

In the large scale, our universe is homogeneous, isotropic and almost flat.
Such a geometrical configuration is well described by the spatially flat
FLRW universe which is characterized by the following line element
\begin{equation}
ds^{2}=-dt^{2}+a^{2}(t)\left( dx^{2}+dy^{2}+dz^{2}\right) ,  \label{metric1}
\end{equation}
where $a(t)$ is the expansion scale factor of the universe. In this
space-time we consider that the main constituents of our universe are a
pressureless matter and a non-canonical scalar field where the matter sector
and the scalar field are interacting with each other through a
non-gravitational interaction.

The Action integral of such a cosmological scenario is given by
\begin{equation}
S=\int d^{4}x\sqrt{-g}\left[ \frac{R}{2k^{2}}-\frac{1}{2}\partial _{\mu
}\phi \partial ^{\mu }\phi -V(\phi )\right] +\mathcal{L}_{m}  \label{action}
\end{equation}%
where $\mathcal{L}_{m}$ denotes the Lagrangian for the matter field terms,
that means pressureless matter and the scalar field. In the background (\ref%
{metric1}), the energy density and the pressure for the scalar field take
the forms
\begin{align}
\rho _{\phi }& =\frac{1}{2}\,\dot{\phi}^{2}+V(\phi ),  \label{energy} \\
p_{\phi }& =\frac{1}{2}\,\dot{\phi}^{2}-V(\phi ),  \label{pressure}
\end{align}%
from where the equation of state parameter $w_{\phi }$ for the scalar field
is defined to be the ratio of its pressure to the energy density, that means,

\begin{equation}
w_{\phi }=\frac{p_{\phi }}{\rho _{\phi }}=\frac{\,\dot{\phi}^{2}-2V\left(
\phi \right) }{\,\dot{\phi}^{2}+2\,V(\phi )}.
\end{equation}

Moreover, we assume that $\rho _{m}$ and $p_{m}$ are respectively the energy
density and pressure of the matter sector. Since we assume the pressureless
matter, thus, we have $p_{m}=0$, and consequently, the equation of state
parameter for this matter sector $w_{m}=0$.

The field equations can be obtained by varying the action with respect to
the metric coefficients $g_{\mu\nu}$ of the space-time as (in the units $%
k^{2}=8\pi G= c= 1 $)

\begin{align}
H^{2}& =\frac{1}{3}(\rho _{m}+\rho _{\phi }),  \label{friedmann} \\
\dot{H}& =-\frac{1}{2}\Bigl(\rho _{m}+\rho _{\phi }+p_{\phi }\Bigr),
\label{Raychaudhuri}
\end{align}%
where an `overdot' represents the cosmic time differentiation, and $H=\dot{a}%
/a$ is the Hubble parameter of the FRLW universe. Furthermore, from the
Bianchi identity we have that $_{\mathrm{eff}}T_{~~;b}^{ab}=0,$ where $_{%
\mathrm{eff}}T^{ab}=_{\phi }T^{ab}+_{m}T^{ab}$. However, because we consider
interaction between the scalar field and the dust fluid, the Bianchi
identity gives the following equations%
\begin{equation}
_{\phi }T_{~~;b}^{ab}+_{m}T_{~~;b}^{ab}=0,
\end{equation}%
or equivalently,%
\begin{align}
\dot{\rho}_{m}+3H\rho _{m}& =Q,  \label{mc} \\
\dot{\rho}_{\phi }+3H(1+w_{\phi })\rho _{\phi }& =-Q,  \label{fc}
\end{align}%
where we have introduced the quantity $Q$ which indicates the rate of energy
exchange between the dark sector. Positive value of $Q$ indicates that there
is an energy transfer from the scalar field $\rho _{\phi }$ to the cold dark
matter $\rho _{m,}~$\ while for $Q<0$, the reverse scenario happens, that
means energy transfer from the cold dark matter to scalar field.

An equivalent way to write the set of equations (\ref{mc}), (\ref{fc}) is
with the use of the ratio $r\left( t\right) =\frac{\rho _{m}}{\rho _{\phi }}$%
, where the equation (\ref{mc}) becomes%
\begin{equation}
\rho _{\phi }\dot{r}-Q\left( 1+r\right) -3Hp_{\phi }r=0.  \label{s12}
\end{equation}

The nature of the interaction function $Q$ is purely unknown and until now
we don't have any device available to derive this function from some
fundamental physical principle. There are various phenomenological
approaches in the literature which have shown that the existence of the
interaction can explain the values of various cosmological parameters during
the late-time acceleration phase of our universe.

Some interaction models which have been proposed in the literature and much
well known to the interaction theory are the linear models, such as $%
Q_{1}=\alpha _{m}H\rho _{m}$ \cite{Amendola:2006dg},~$Q_{2}=\alpha _{\phi
}H\rho_{\phi }$ \cite{Pavon:2007gt} and \ $Q_{3}=$ $\alpha \left( H\rho
_{m}+H\rho _{\phi }\right) ~$ \cite{Quartin:2008px}. While only a few
nonlinear models have been proposed and investigated in the literature \cite%
{Chimento:2009hj, Arevalo:2011hh, Yang:2017zjs}.

In this work we shall study the evolution of the field equations (\ref%
{friedmann}), (\ref{Raychaudhuri}), (\ref{mc}) and\ (\ref{fc}) for some
nonlinear interacting models of the form $Q\left( \rho _{m},\rho _{\phi
}\right) $. Such an analysis provides us with the information for different
phases of the universe provided by the field equations and also the
stability of these phases as well. In order to perform our analysis we
prefer to work with the dimensionless variables, and more specifically, we
select to work with the so-called $H-$normalization as followed in \cite%
{Copeland:2006wr}.

\section{Construction of the Dynamical system}

\label{formation}

We continue with the introduction of the dimensionless variables \cite%
{Copeland:2006wr}

\begin{equation}
x=\frac{\dot{\phi}}{\sqrt{6}H},~~~y=\frac{\sqrt{V(\phi )}}{\sqrt{3}\,H},
\label{new-variables}
\end{equation}%
in the $H$-normalization. Moreover, we assume that the new independent
variable is the lapse time $N=\ln a$, which is also called the $e$-folding
parameter.

In the new variables the gravitational field equations (\ref{friedmann}), (%
\ref{Raychaudhuri}), (\ref{fc})~and (\ref{s12}) reduce to the following
autonomous system of algebraic-differential system%
\begin{align}
\frac{dx}{dN}& =-3x+\sqrt{\frac{3}{2}}\,\lambda y^{2}+\frac{3x}{2}\left(
\left( 2+r\right) x^{2}+ry^{2}\right) -\left( 6x\right) ^{-1}\bar{Q},
\label{de1} \\
\frac{dy}{dN}& =-3\sqrt{2}\,\lambda xy\,+3\sqrt{3}y\left( \left( 2+r\right)
x^{2}+ry^{2}\right) ,  \label{de2} \\
\frac{d\lambda }{dN}& =-\sqrt{6}\,x\,\lambda ^{2}\left( \Gamma \left(
\lambda \right) -1\right) ,  \label{de4} \\
\frac{dr}{dN}& =3\left( x^{2}+y^{2}\right) ^{-1}\left( \left( 1+r\right)
\bar{Q}+9r\left( x^{2}-y^{2}\right) \right) ,  \label{de4a}
\end{align}%
with the algebraic constraint%
\begin{equation}
1-\left( 1+r\right) \left( x^{2}+y^{2}\right) =0,  \label{de1aa}
\end{equation}%
in which%
\begin{equation}
\lambda =-\frac{V_{,\phi }}{V},~\bar{Q}=\frac{Q}{\,H^{3}},~\text{and }%
~\Gamma =\frac{VV_{,\phi \phi }}{(V_{,\phi })^{2}}.  \label{de5}
\end{equation}

Furthermore, the first Friedmann equation (\ref{friedmann}) in the
dimensionless variables becomes%
\begin{equation}
\Omega _{m}=1-x^{2}-y^{2},  \label{de5a}
\end{equation}%
where $\Omega _{m}=\frac{1}{3}\rho _{m}H^{-2}$, is the dimensionless density
parameter for the matter sector. The cosmological parameters for the scalar
field, namely the equation of state parameter $w_{\phi }$ and the
dimensionless density parameter $\Omega _{\phi }$ in the new variables can
be written as follows
\begin{equation}
w_{\phi }= \frac{x^{2}-y^{2}}{x^{2}+y^{2}},~~\Omega _{\phi }=\frac{\rho
_{\phi }}{3H^{2}}= x^{2}+y^{2}  \label{de.6}
\end{equation}%
while the total equation of state parameter is
\begin{equation}
w_{tot}=\frac{p_{\phi }}{\rho _{m}+\rho _{\phi }}=\,x^{2}-y^{2}.
\label{de.7}
\end{equation}

Finally, one can calculate the deceleration parameter that assumes the
following expression
\begin{equation}
q=-1-\frac{\dot{H}}{H^{2}}=\frac{1}{2}(1+3\left( \,x^{2}-y^{2}\right) ).
\label{de.8}
\end{equation}

As far as concerns, for the solution of the scale factor, at any point $%
\left( x_{0},y_{0},\lambda \right) ,$ we have that $w_{tot}=const.;$ hence
from (\ref{Raychaudhuri}) it follows that $a\left( t\right) \propto t^{\frac{%
2}{3\left( 1+w_{tot}\right) }}$ for $w_{tot}\neq -1$ and $a\left( t\right)
=a_{0}e^{H_{0}t}$ for $w_{tot}=-1$, where the latter corresponds to the de
Sitter points.

Due to the energy condition $0\leq \Omega _{m}\leq 1$, we have two different
conditions which depends on the nature of the scalar field. If the action
integral for the scalar field is that of the quintessence, i.e. $\varepsilon
=1$, it follows that $x^{2}+y^{2}\leq 1$. Which means that the parameters $%
\left\{ x,y,z\right\} $ take values inside a unit sphere. However from the
definition of the dimensionless variables (\ref{new-variables}) we have $%
y\in \left[ 0,1\right] $ and $\left\{ x\right\} \in \left[ -1,1\right] $.

However, because we shall work with interactions of the form $\bar{Q}\left(
r,x,y\right) $, the limit where $\Omega _{\phi }=x^{2}+y^{2}\rightarrow 0$,
corresponds to $r\rightarrow +\infty ,$ consequently to the matter dominated
era.

As far as concerns the nonlinear interaction models of our analysis we
consider the following
\begin{equation}
\bar{Q}_{A}=Q_{0}\frac{r}{x^{2}}~~,\bar{Q}_{B}=Q_{0}\frac{x^{2}}{r}~~,~\bar{Q%
}_{C}=Q_{0}\frac{y^{2}}{x^{2}+y^{2}}r^{2},  \label{models}
\end{equation}%
where $Q_{0}$ is a constant and it is the interaction parameter or the
coupling parameter. The parameter could describe the strength of interaction
(from its magnitude) and the direction of energy flow between the dark
sectors (through its sign).

In terms of the energy densities $\rho _{m},~\rho _{\phi }$ the interactions
in (\ref{models}) are expressed as
\begin{equation}
\bar{Q}_{A}\simeq \dot{\phi}^{-2}\frac{\rho _{m}}{\rho _{\phi }}~,~\bar{Q}%
_{B}\simeq \dot{\phi}^{2}\frac{\rho _{\phi }}{\rho _{m}}~,~\bar{Q}\simeq
V\left( \phi \right) \frac{\left( \rho _{m}\right) ^{2}}{\left( \rho _{\phi
}\right) ^{3}}
\end{equation}

With the use of the constraint (\ref{de1aa}) we can reduce the dynamical
system (\ref{de1})-(\ref{de4a}) to a three dimensional system by replacing $%
r=\left( x^{2}+y^{2}\right) ^{-1}-1$; while when $\lambda =const.$, the
dynamical system is reduced to a two dimensional system because equation (%
\ref{de4}) is identically satisfied. The only potential function $V\left(
\phi \right) $ where $\lambda =const.$, always, is the exponential potential
$V\left( \phi \right) =V_{0}e^{-\lambda \phi }$ \cite{Copeland:1997et}.

\section{Critical points and stability}

\label{sec-critical}

Before we proceed with the specific interactions let us consider the most
general scenario $\bar{Q}=\bar{Q}\left( r,x,y\right) $. By replacing this
interaction into equation (\ref{de1}), the critical points for the
two-dimensional dynamical system (\ref{de1}), (\ref{de2}) in which
\begin{equation}
r=\frac{1-x^{2}-y^{2}}{x^{2}+y^{2}},
\end{equation}%
can be categorized into two families. Family (A) consists of the points with
$y=0$, and Family (B) consists of the points with $y\neq 0$; more
specifically,
\begin{equation}
y^{2}=1+x^{2}-\sqrt{\frac{2}{3}}\lambda x.
\end{equation}%
These two families follow from the solution of the algebraic equation
\begin{equation}
\left[ \left( 2+r\right) x^{2}+ry^{2}-\sqrt{\frac{2}{3}}\,\lambda xy\right]
y=0.
\end{equation}

\begin{figure}[tbp]
\includegraphics[height=5cm]{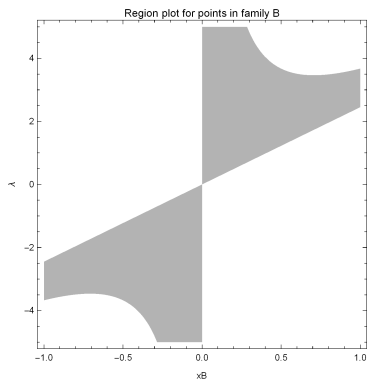}
\caption{Region plot in the space $\left\{ x_{B},\protect\lambda \right\} $,
where the points of Family B exist according to the constraint (\protect\ref%
{ss.03}).}
\label{reglambda}
\end{figure}

As far as concerns, the total number of points corresponding to each family,
depends on the prescribed interaction function $\bar{Q}\left( r,x,y\right) $%
, and also on how many real solutions are admitted by the algebraic equation%
\begin{equation}
-3x+\sqrt{\frac{3}{2}}\,\lambda y^{2}+\frac{3x}{2}\left( \left( 2+r\right)
x^{2}+ry^{2}\right) -\left( 6x\right) ^{-1}\bar{Q}\left( r,x,y\right) =0~,
\end{equation}%
for $x\in \left[ -1,1\right] $. Moreover, for the points of Family A the
physical parameters are simplified as%
\begin{equation}
\Omega _{m}\left( x_{A}\right) =1-x_{A}^{2}~,~w_{\phi }\left( A\right)
=1~,~w_{tot}=\,x_{A}^{2}~,
\end{equation}%
which means that the scalar field acts as a stiff fluid, while because of
the interaction $x_{A}$ can be different from zero, that means the dust
fluid can contribute to the universe. For the points of Family B, the
physical parameters are written in terms of $x_{B}$ as follows%
\begin{equation}
\Omega _{m}\left( x_{B},\lambda \right) =\frac{\sqrt{6}}{3}x_{B}\left(
\lambda -\sqrt{6}x_{B}\right) ~,~  \label{ss.01}
\end{equation}%
\begin{equation}
w_{\phi }\left( x_{B},\lambda \right) =\frac{\sqrt{6}\lambda x_{B}-3}{3+%
\sqrt{6}x_{B}\left( \lambda -\sqrt{6}x\right) },
\end{equation}%
and
\begin{equation}
w_{tot}\left( x_{B},\lambda \right) =-1+\sqrt{\frac{2}{3}}\lambda x_{B}.
\end{equation}%
At this point, from (\ref{ss.01}) one can find the constraint on $\lambda $
(recall that the constraint on $\Omega _{m}$ is, $0\leq \Omega _{m}\leq 1$)
as
\begin{equation}
\sqrt{6}\left\vert x_{B}\right\vert \leq \lambda \leq \frac{3+6\left\vert
x_{B}\right\vert ^{2}}{\sqrt{6}\left\vert x_{B}\right\vert }.  \label{ss.03}
\end{equation}%
In Fig. \ref{reglambda} the area in the space $\left\{ x_{B},\lambda
\right\} $ is presented where the points of Family B exist.

We continue our analysis by presenting the physical parameters for the
critical points of Family A and Family B, and also the region where the
critical points are stable. Special cases of study are given.

\subsection{Interaction $\bar{Q}_{A}$}

For the nonlinear interaction $\bar{Q}_{A}$, the critical points are
\begin{equation}
P_{A}\left( \bar{Q}_{A}\right) =\left( \pm \left( -\frac{Q_{0}}{9}\right) ^{%
\frac{1}{6}},0\right)
\end{equation}%
\begin{equation}
P_{B}\left( \bar{Q}_{A}\right) =\left( x_{B},\sqrt{1+x_{B}^{2}-\sqrt{\frac{2%
}{3}}\lambda x_{B}}\right)
\end{equation}%
where $x_{B}$ is a solution of the fourth-order algebraic equation%
\begin{equation}
Q_{0}=3x_{B}^{2}\left( 3-2x_{B}\left( \sqrt{6}\lambda \left(
1+x_{B}^{2}\right) -\left( 3+\lambda ^{2}\right) x_{B}\right) \right)
\label{ss0001}
\end{equation}

From the algebraic expressions of $P_{A}\left( \bar{Q}_{A}\right) $ and $%
P_{B}\left( \bar{Q}_{B}\right) $, one can infer that Family A consists of
two critical points, while Family B is composed by 0, 2 or 4 critical points.

\subsubsection{Critical Points of Family A}

Points $P_{A}\left( \bar{Q}_{A}\right) $ are real only when $Q_{0}<0$, while
exits for values of $\left\vert Q_{0}\right\vert $ in the range
\begin{equation}
0<\left\vert Q_{0}\right\vert \leq 9,
\end{equation}%
while the total equation of state parameter $w_{tot}$ is constrained by
\qquad\
\begin{equation}
0<w_{tot}\leq 1.
\end{equation}

As far as concerns the stability of the critical points $P_{A}\left( \bar{Q}%
_{A}\right) $ we derive the eigenvalues
\begin{equation}
e_{1}\left( P_{A}\left( \bar{Q}_{A}\right) \right) =6~,~e_{2}\left(
P_{A}\left( \bar{Q}_{A}\right) \right) =0
\end{equation}%
from which one can easily infer that the critical points $P_{A}\left( \bar{Q}%
_{A}\right) ~$are unstable in nature.

\subsubsection{Critical Points of Family B}

\begin{figure}[tbp]
\includegraphics[width=0.4\textwidth]{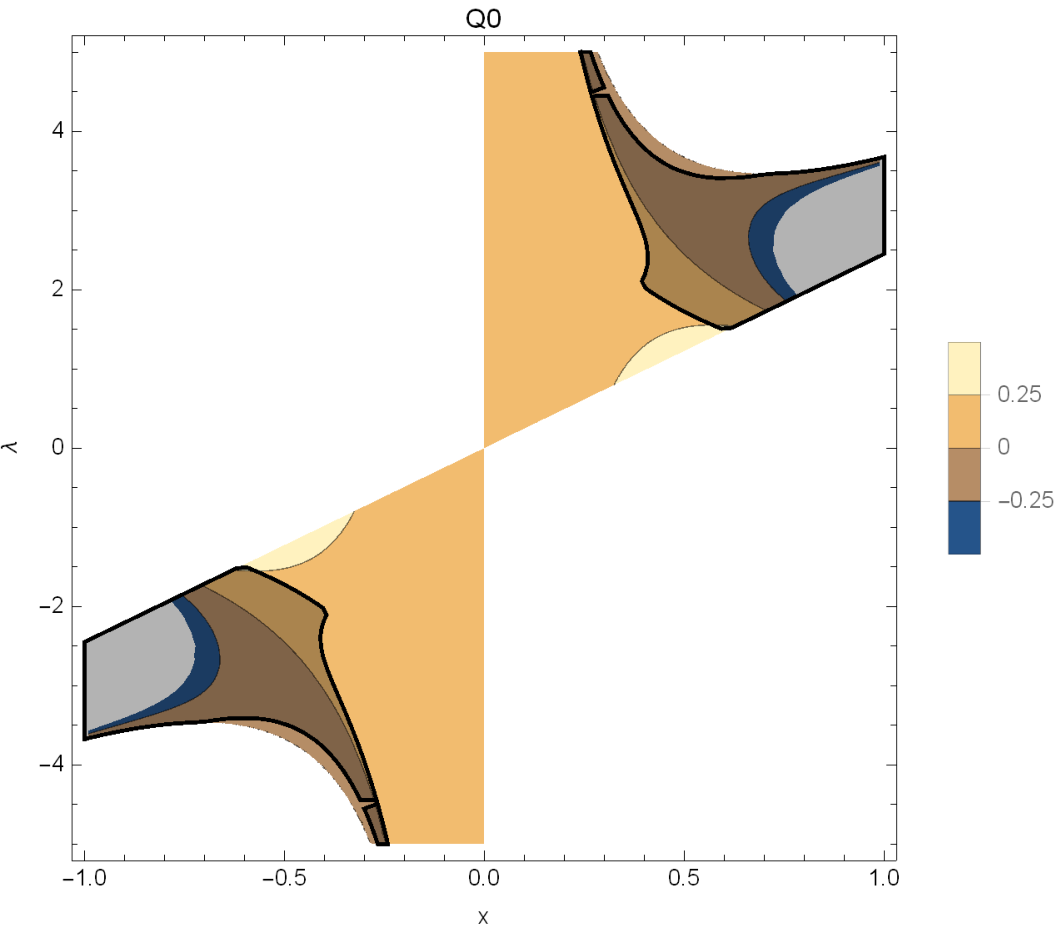} \includegraphics[width=0.4%
\textwidth]{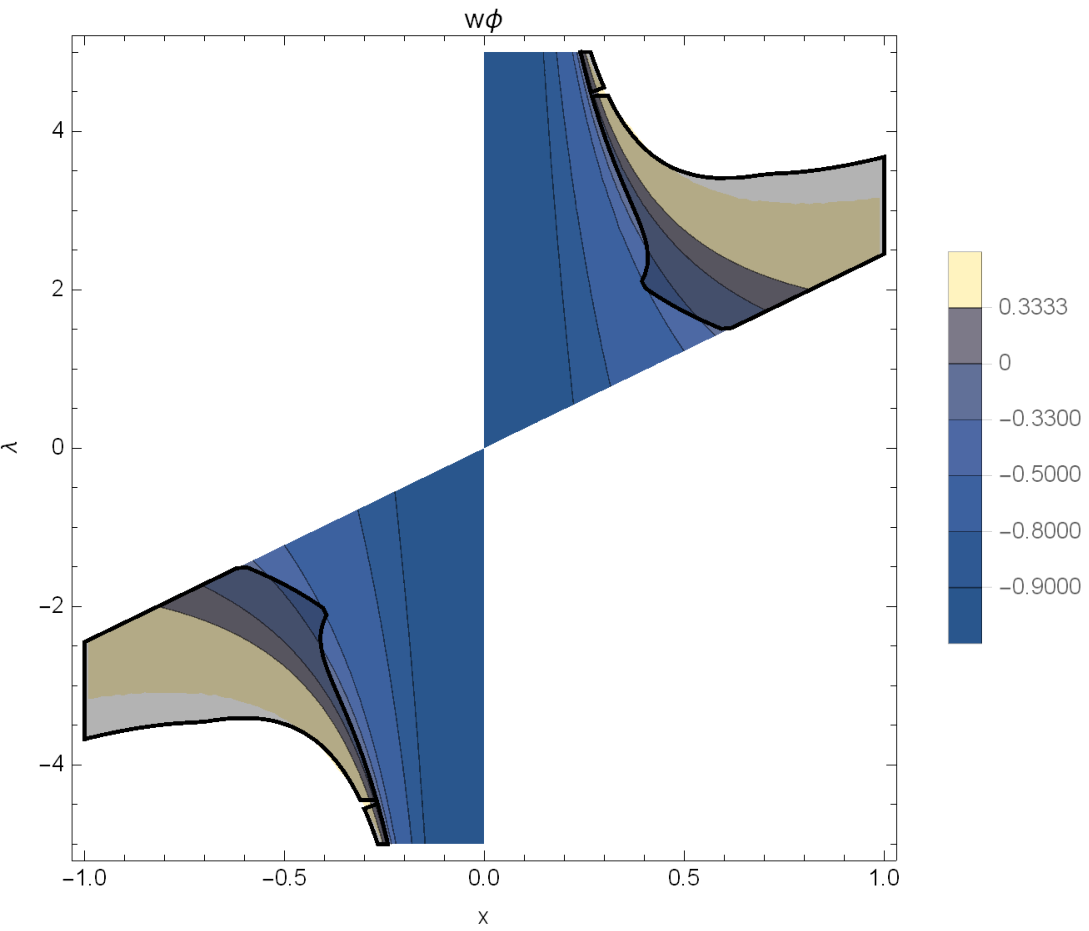} \newline
\includegraphics[width=0.4\textwidth]{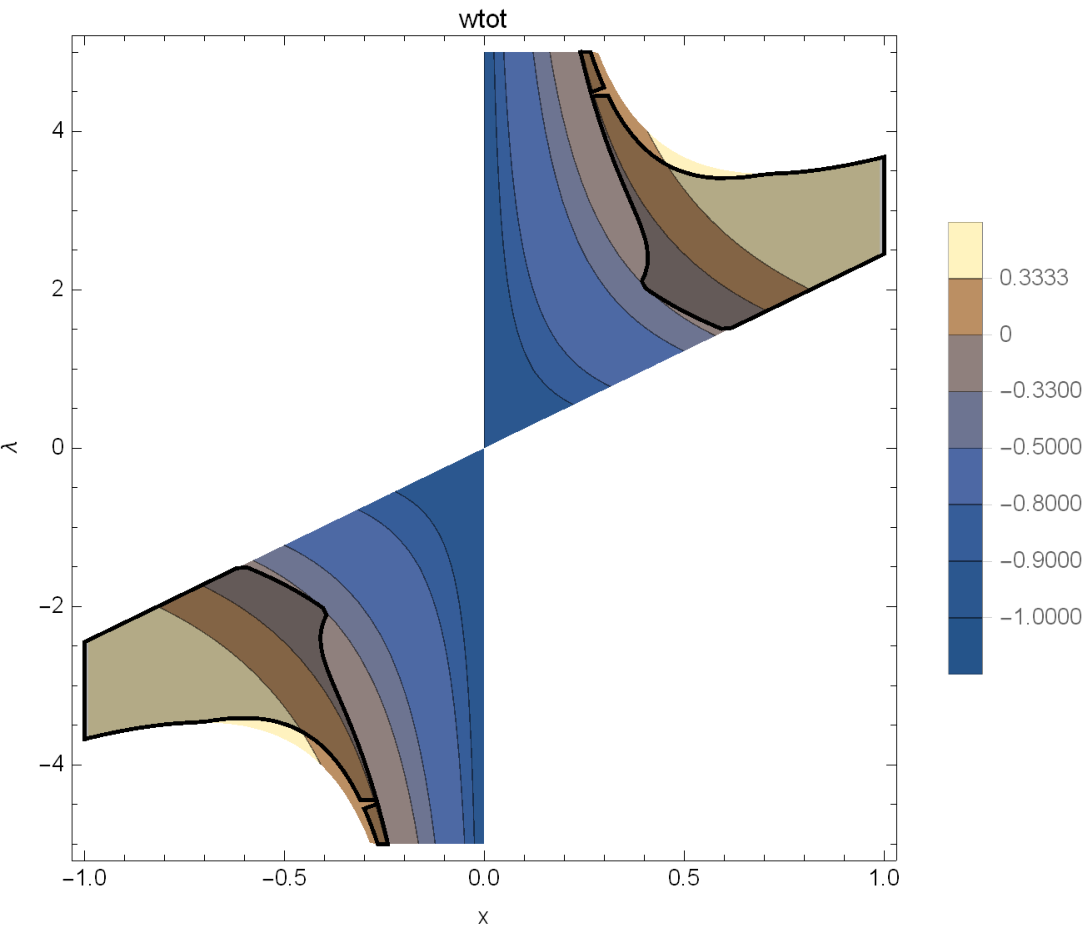} \includegraphics[width=0.4%
\textwidth]{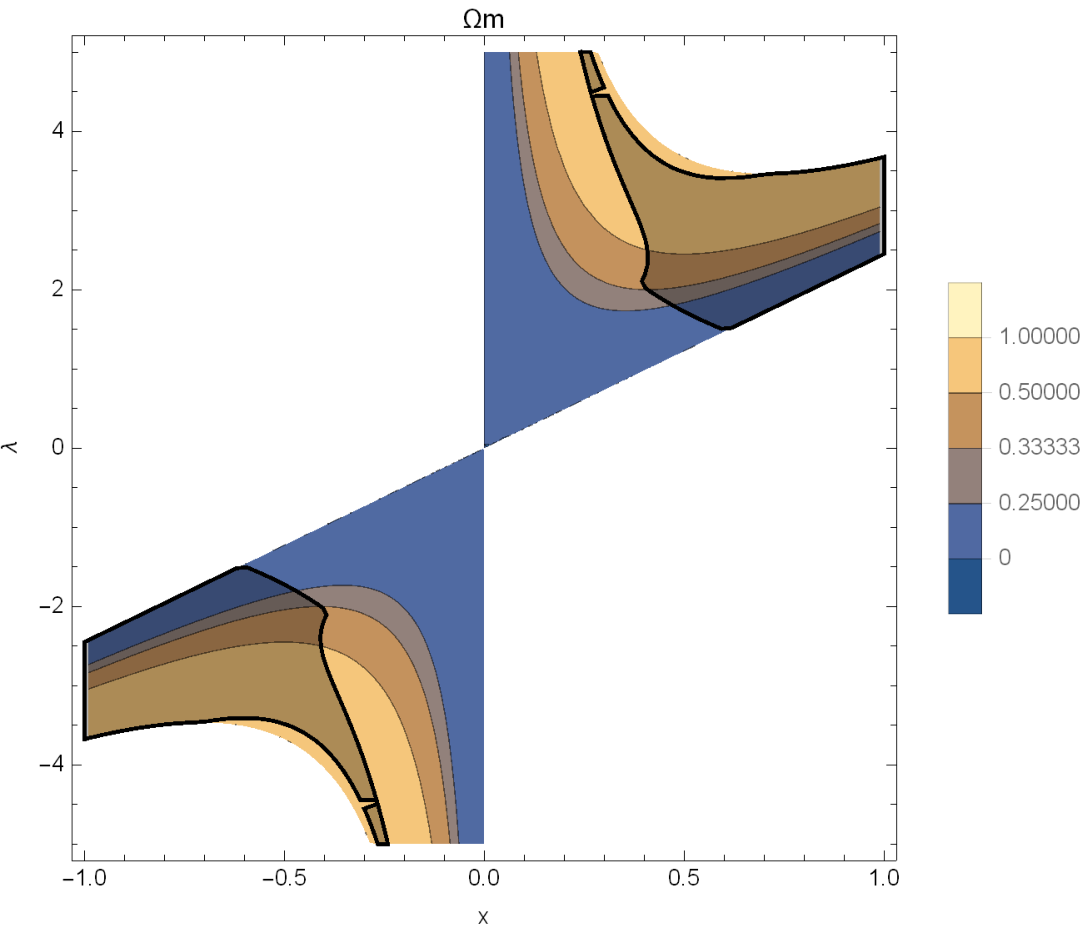}
\caption{Region plot in the space $\left\{ x_{B},\protect\lambda \right\} $
for the interaction constant $Q_{0}$, and also the physical parameters $%
\Omega_{m}$, $w_{\protect\phi }$, $w_{tot}$ for the points which belong to
Family B of the interaction $\bar{Q}_{A}$. The shaded regions define the
areas where the critical points are stable. }
\label{modelapb}
\end{figure}

Because of the nonlinearity of the algebraic equation (\ref{ss0001}) we are
unable to get the exact expressions for the physical parameters and also the
stability of the points as well. Thus, we proceed with the numerical
solution of (\ref{ss0001}) in order to investigate the cosmological
parameters. Following this, in Fig. \ref{modelapb} the contour plots for the
physical parameters $\Omega _{m}\left( P_{B}\left( \bar{Q}_{A}\right)
\right) ,~$ $w_{\phi }\left( P_{B}\left( \bar{Q}_{A}\right) \right) $ and $%
w_{tot}\left( P_{B}\left( \bar{Q}_{A}\right) \right) $ are presented, and
also the specific values of the surface (\ref{ss0001}) in the plane $%
x_{B}-\lambda $, using the constraint in (\ref{ss.03}). In addition, the
shaded areas in Fig. \ref{modelapb} define the region of the parameter $%
\left\{ x_{B},\lambda \right\} $, where the critical points are stable.

From the plots in Fig. \ref{modelapb} we can infer that there exist regions
of the parameters $\left\{ x_{B},\lambda \right\} $ where the stable
critical points cannot describe an accelerated universe, since we always
have $-\frac{1}{3}<w_{tot}$, and there we have $\Omega _{m}\left(
x_{B},\lambda \right) \neq 0$. \ On the other hand, the de Sitter universe
can be described only as an unstable solution for the current interaction
model.

We now proceed by considering a specific value for the parameter $\lambda $.

\bigskip

\paragraph{Special case $\protect\lambda =2:$}

\begin{figure}[tbp]
\includegraphics[width=0.4\textwidth]{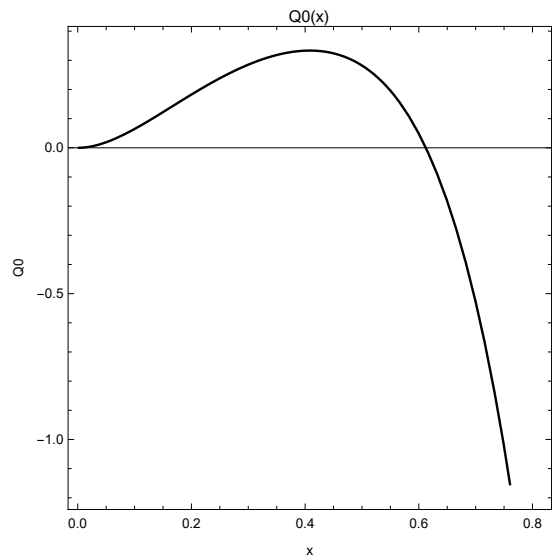}
\caption{The figure describes the numerical solution for the polynomial
equation (\protect\ref{pf.01}).}
\label{sf1aaa}
\end{figure}

Let us now consider a specific value of $\lambda =2$. One can clearly see
that the points $\bar{P}_{A}\left( Q_{A}\right) $ do not change since there
is no $\lambda $ dependence, so we focus on points $P_{B}\left( Q_{B}\right)
$. For that specific value of the parameter $\lambda $, the algebraic
equation (\ref{ss0001}) is simplified as follows%
\begin{equation}
Q_{0}=3x_{B}^{2}\left( 3-2x_{B}\left( 2\sqrt{6}\left( 1+x_{B}^{2}\right)
-7x_{B}\right) \right)  \label{pf.01}
\end{equation}%
while from (\ref{ss.03}) parameter $x_{B}$ is constrained to be $x_{B}\in (0,%
\sqrt{2/3}~]$. In that range, from Fig. \ref{sf1aaa} it is clear that there
exists two real points which correspond to the Family B.

\begin{figure}[tbp]
\includegraphics[height=5cm]{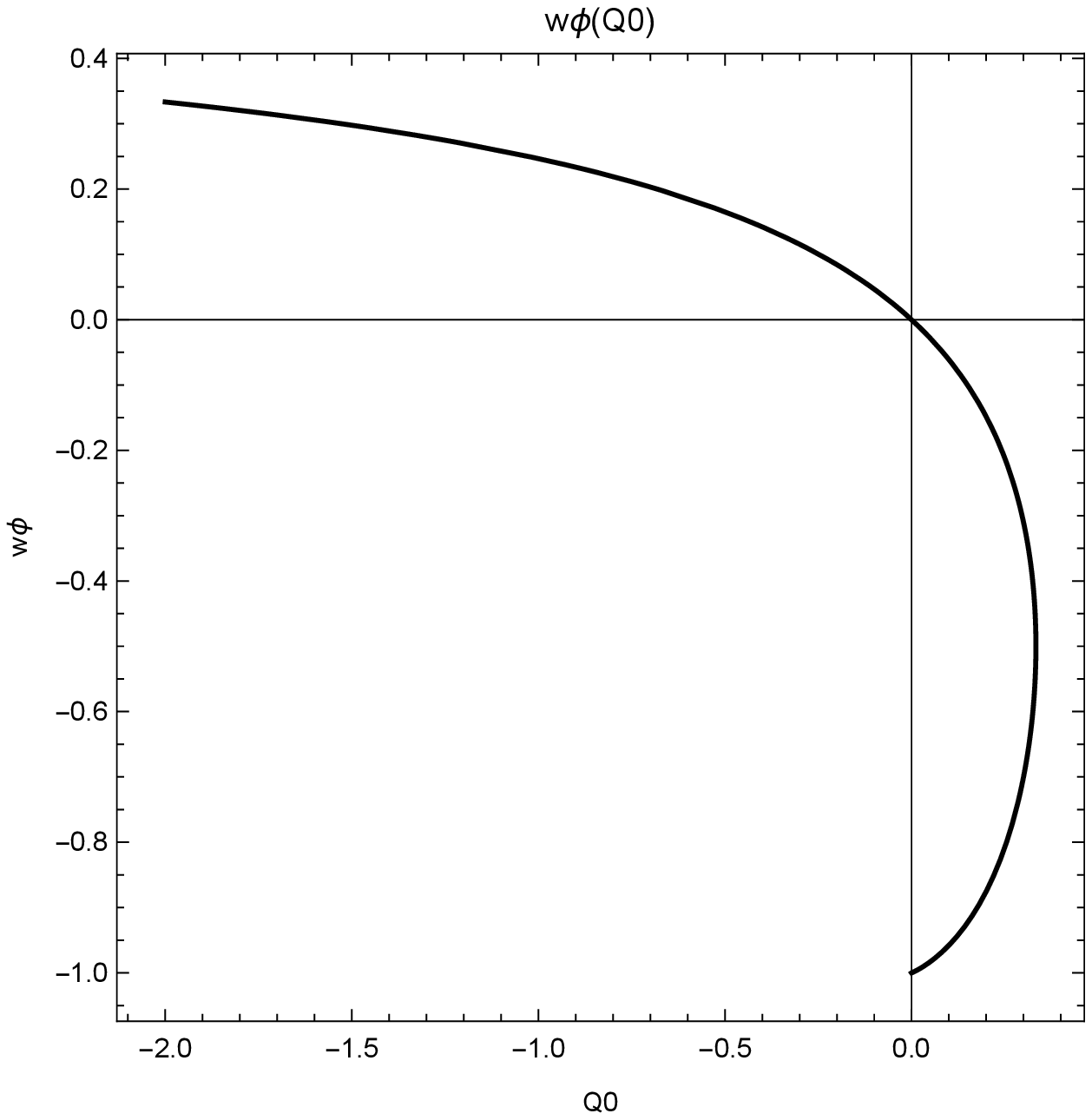}\centering%
\includegraphics[height=5cm]{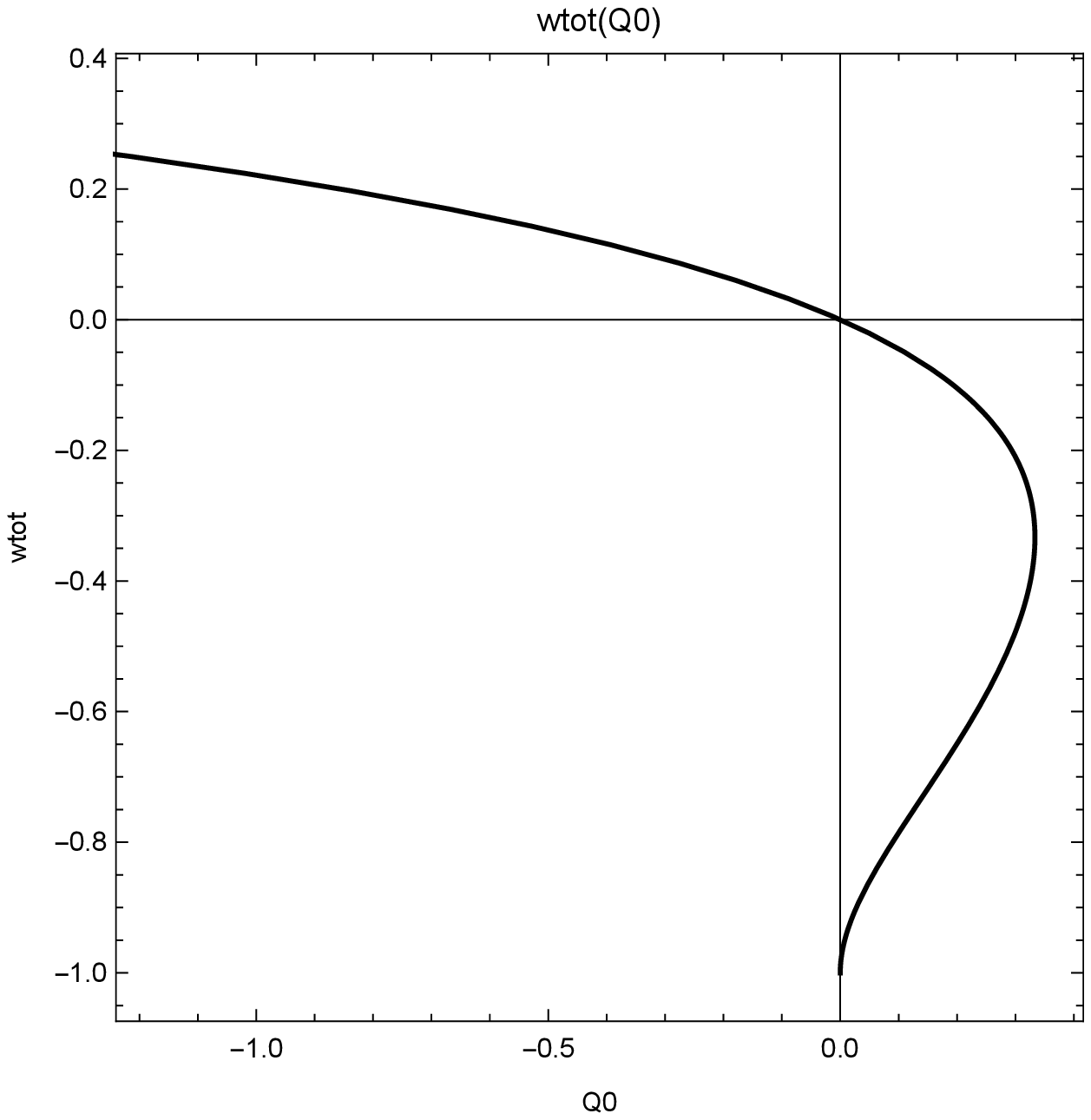}\centering
\newline
\includegraphics[height=5cm]{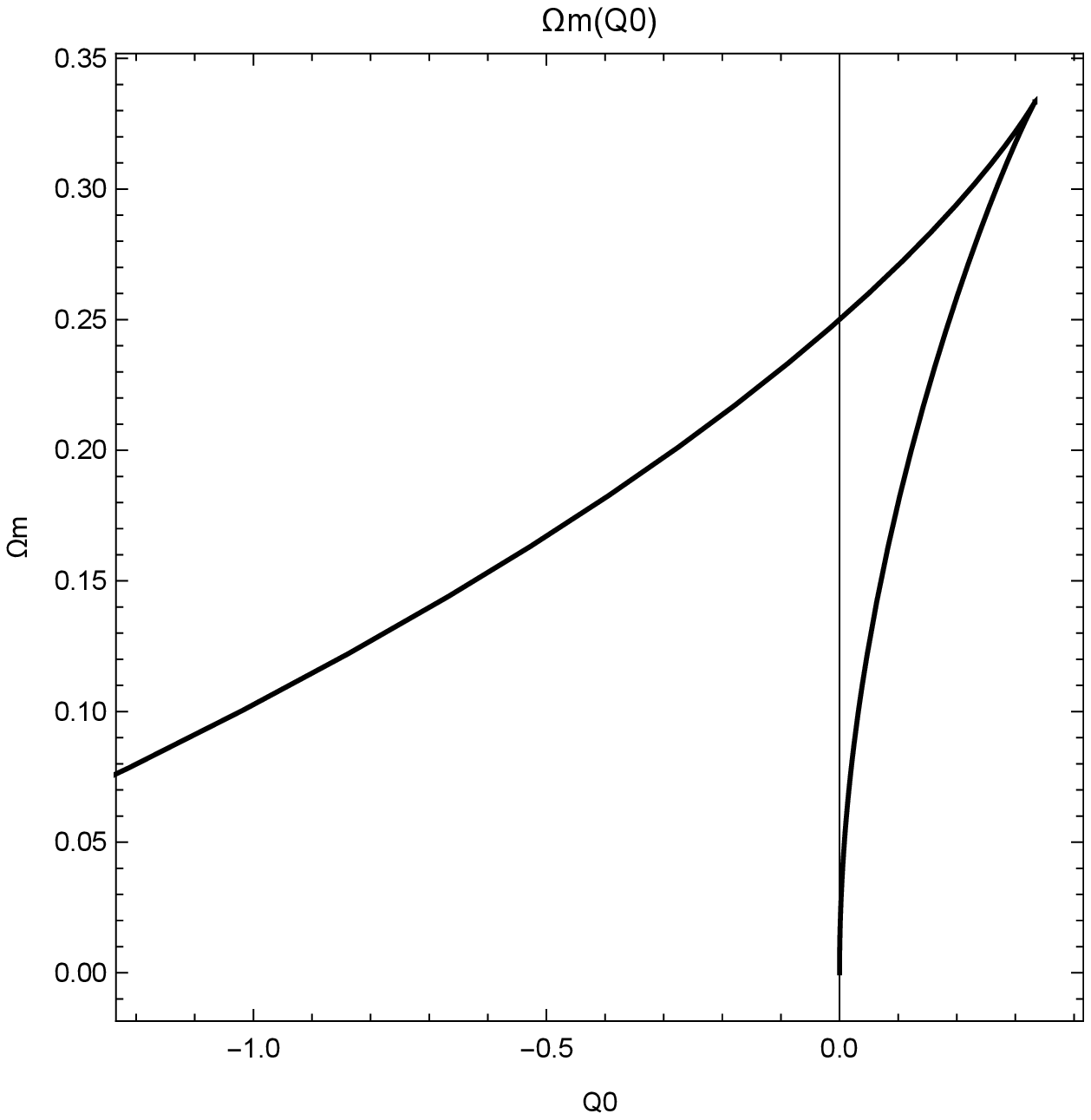}\centering%
\includegraphics[height=5cm]{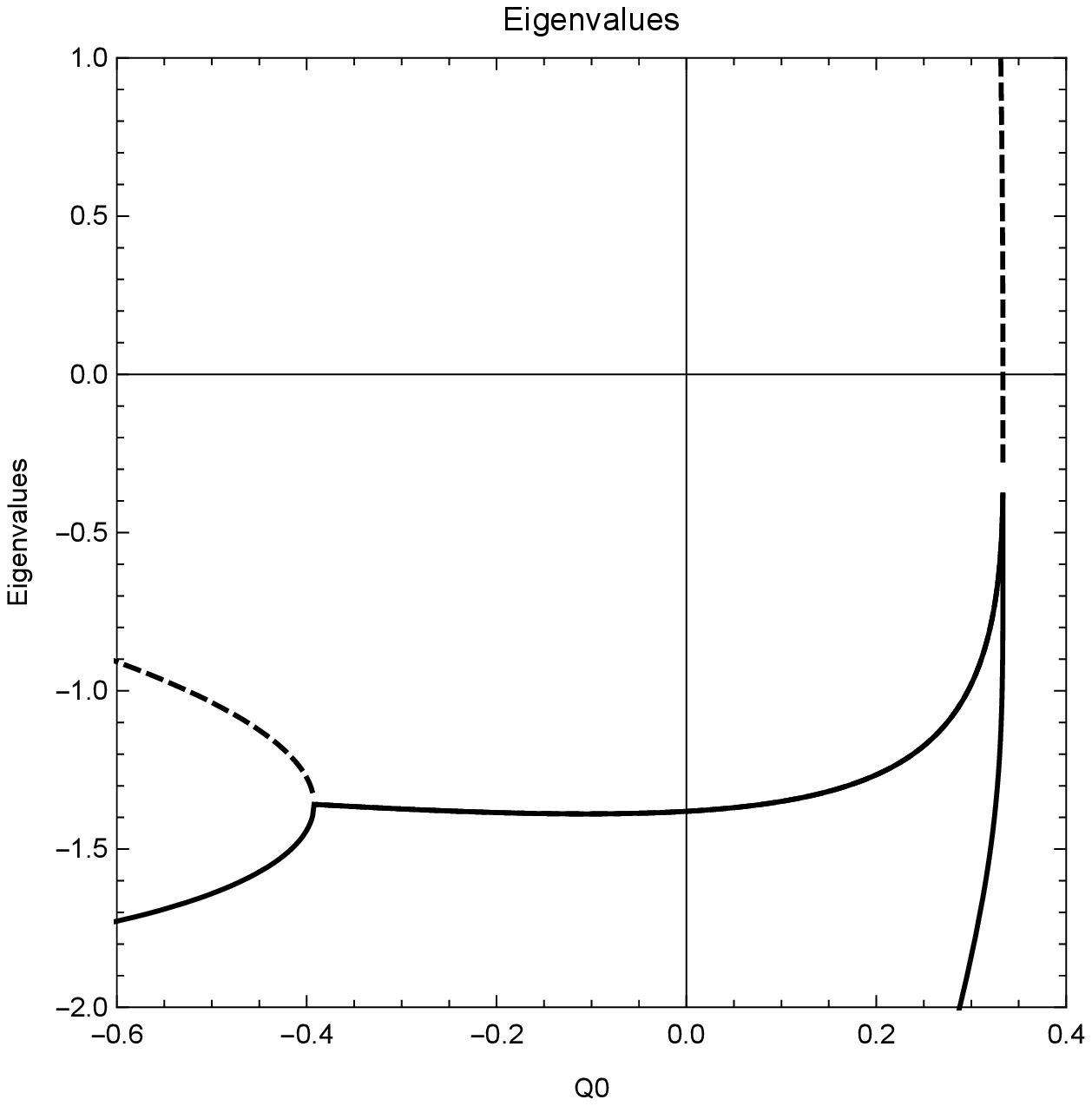}\centering
\caption{Evolution of the physical parameters $w_{\protect\phi }\left(
Q_{0}\right) ,~w_{tot}\left( Q_{0}\right) $ and $\Omega _{m}\left(
Q_{0}\right) $ for the interaction model $Q_{A}$ and $\protect\lambda =2$ at
the critical points of Family B. The evolution of the eigenvalues of the
linearized system is given where we observe that the points are stable for $%
-2<Q_{0}<\frac{1}{3}$}
\label{qal1}
\end{figure}

In Fig. \ref{qal1} the qualitative evolution of the physical parameters $%
w_{\phi },~w_{tot}$ and $\Omega _{m}$ in terms of $Q_{0}$ are given, and
also the evolution of the eigenvalues of the linearized system near the
critical points. We conclude that for that specific value of $\lambda =2$,
the critical points are stable for $\frac{\sqrt{6}}{6}<x_{B}<\sqrt{\frac{2}{3%
}}~$which means%
\begin{equation}
-2<Q_{0}<\frac{1}{3}
\end{equation}%
while in that space of variables%
\begin{equation}
-\frac{1}{3}<w_{tot}<\frac{1}{3}~,~-\frac{1}{2}<w_{\phi }<\frac{1}{3}
\end{equation}%
and%
\begin{equation}
0<\Omega _{m}<\frac{1}{3}.
\end{equation}%
That means the present accelerated solution is not admissible for this case.

\subsection{Interaction $\bar{Q}_{B}$}

Interaction $\bar{Q}_{B}$ is the reverse function of interaction $\bar{Q}%
_{A} $, that is, $\bar{Q}_{B}\simeq \left( \bar{Q}_{A}\right) ^{-1}$. For
that interaction we find the following real critical points
\begin{equation}
P_{A}\left( \bar{Q}_{B}\right) =\left( \pm \frac{\sqrt{18+Q_{0}-\sqrt{%
Q_{0}\left( 36+Q_{0}\right) }}}{3\sqrt{2}},0\right)
\end{equation}%
\begin{equation}
P_{B}\left( \bar{Q}_{B}\right) =\left( x_{B},\sqrt{1+x_{B}^{2}-\sqrt{\frac{2%
}{3}}\lambda x_{B}}\right)
\end{equation}%
where $x_{B}$ is given by the following algebraic equation%
\begin{equation}
Q_{0}=6\left( 3-\sqrt{6}\lambda x_{B}+\lambda ^{2}-9\left( 3+6x_{B}^{2}-%
\sqrt{6}\lambda x_{B}\right) ^{-1}\right) .  \label{polqb}
\end{equation}%
Hence, Family A admits two critical points, while Family B admits 1 or 3
real critical points.

\subsubsection{Critical Points of Family A}

The energy density for the dust fluid for the points of Family A is
calculated to be%
\begin{equation}
\Omega _{m}\left( P_{A}\left( \bar{Q}_{B}\right) \right) =\frac{1}{18}\left(
\sqrt{Q_{0}\left( 36+Q_{0}\right) }-Q_{0}\right)
\end{equation}%
from where we can infer that $Q_{0}\geq 0$ in order $\Omega _{m}$ to be
bounded.

Moreover, for points $P_{A}\left( \bar{Q}_{B}\right) $ we calculate the two
eigenvalues of the linearized system given by
\begin{equation}
e_{1}\left( P_{A}\left( \bar{Q}_{B}\right) \right) =\frac{1}{6}\left(
36+Q_{0}-\sqrt{Q_{0}\left( 36+Q_{0}\right) }\right) ~\text{and }e_{2}\left(
P_{A}\left( \bar{Q}_{B}\right) \right) =0.
\end{equation}

Therefore, for $Q_{0}>0\,\ ${we find that always ${Re}\left[ e_{1}\left( \
P_{A}\left( \ \bar{Q}_{B}\right) \ \right) \right] >0$,} consequently we
conclude that points $P_{A}\left( \bar{Q}_{B}\right) $ are unstable.

\subsubsection{Critical Points of Family B}

\begin{figure}[tbp]
\includegraphics[height=5cm]{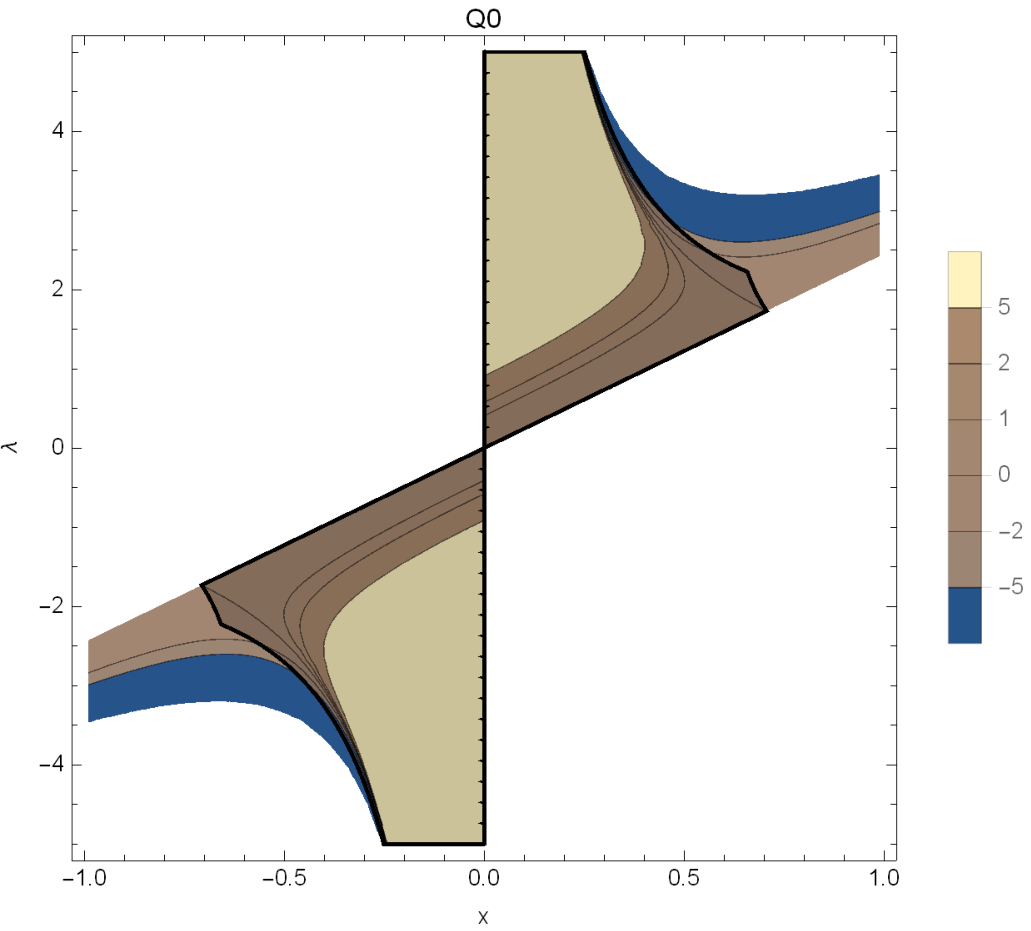}\centering%
\includegraphics[height=5cm]{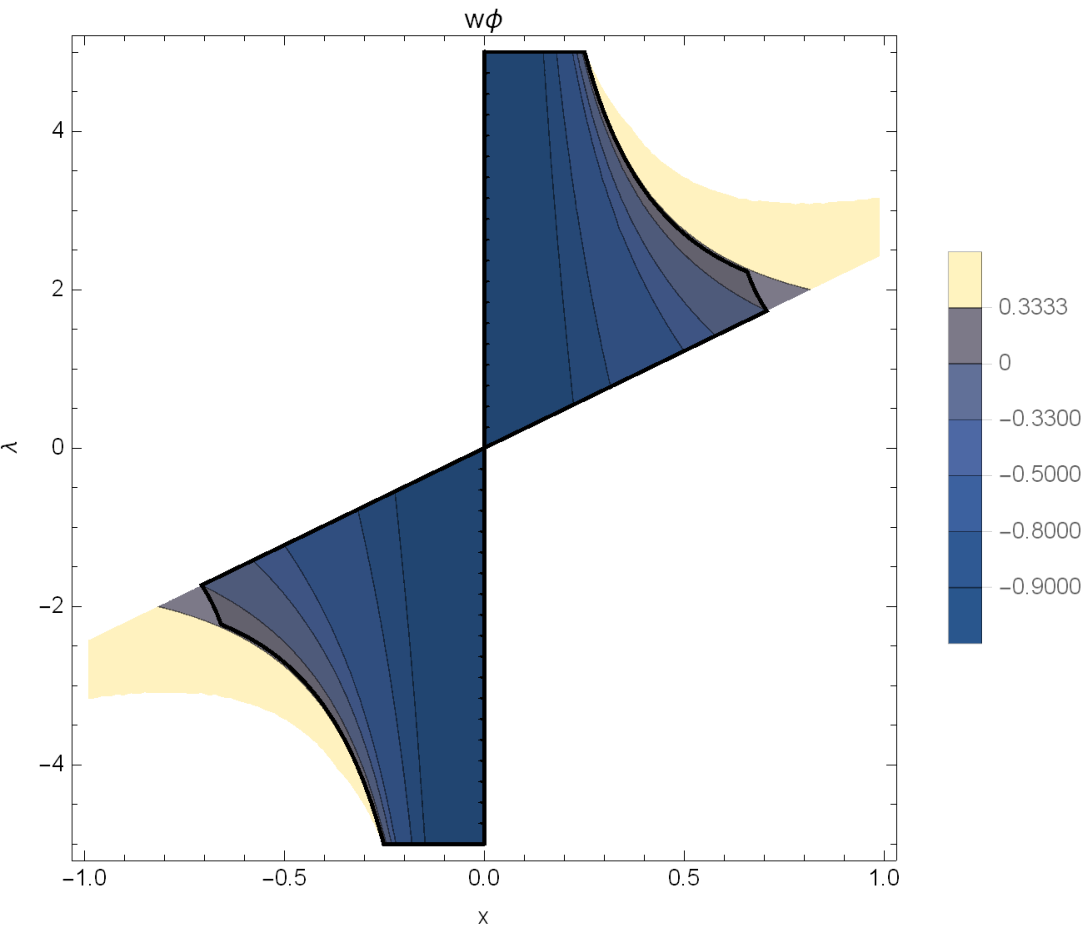}\centering
\newline
\includegraphics[height=5cm]{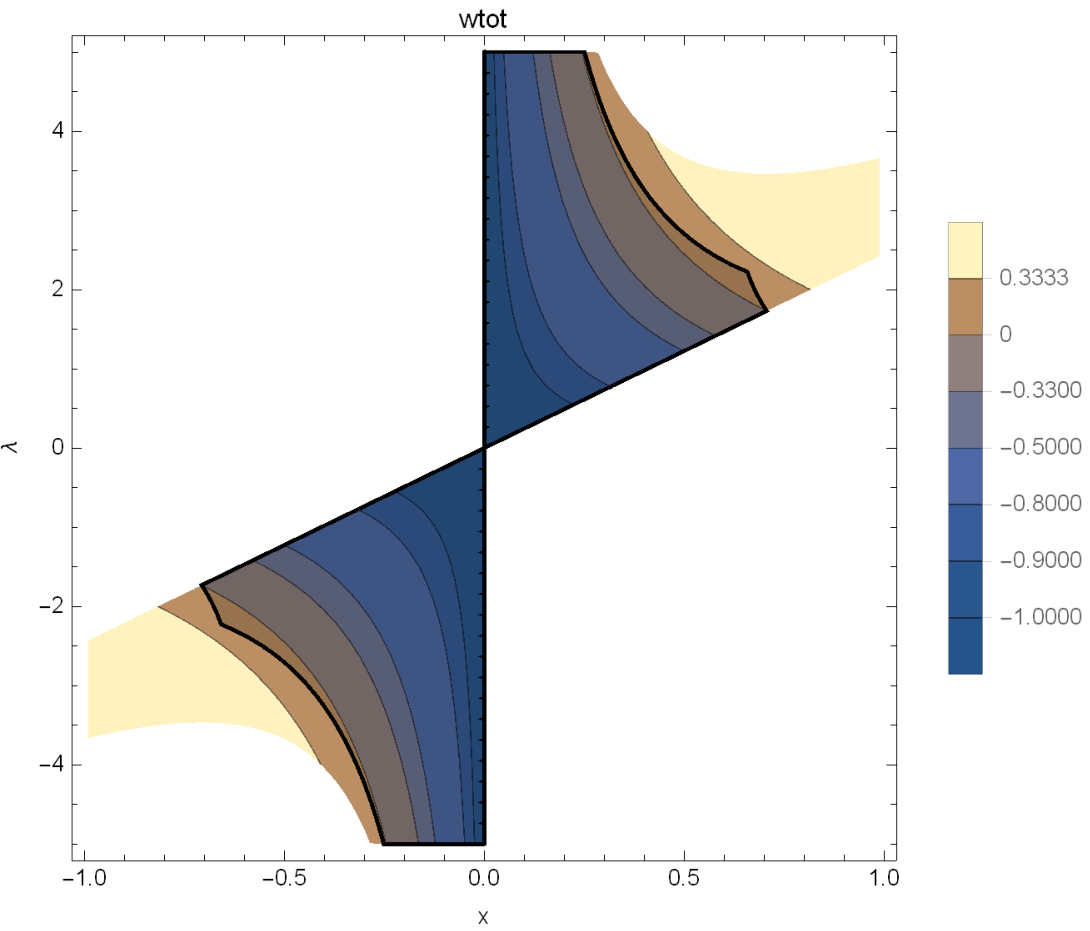}\centering%
\includegraphics[height=5cm]{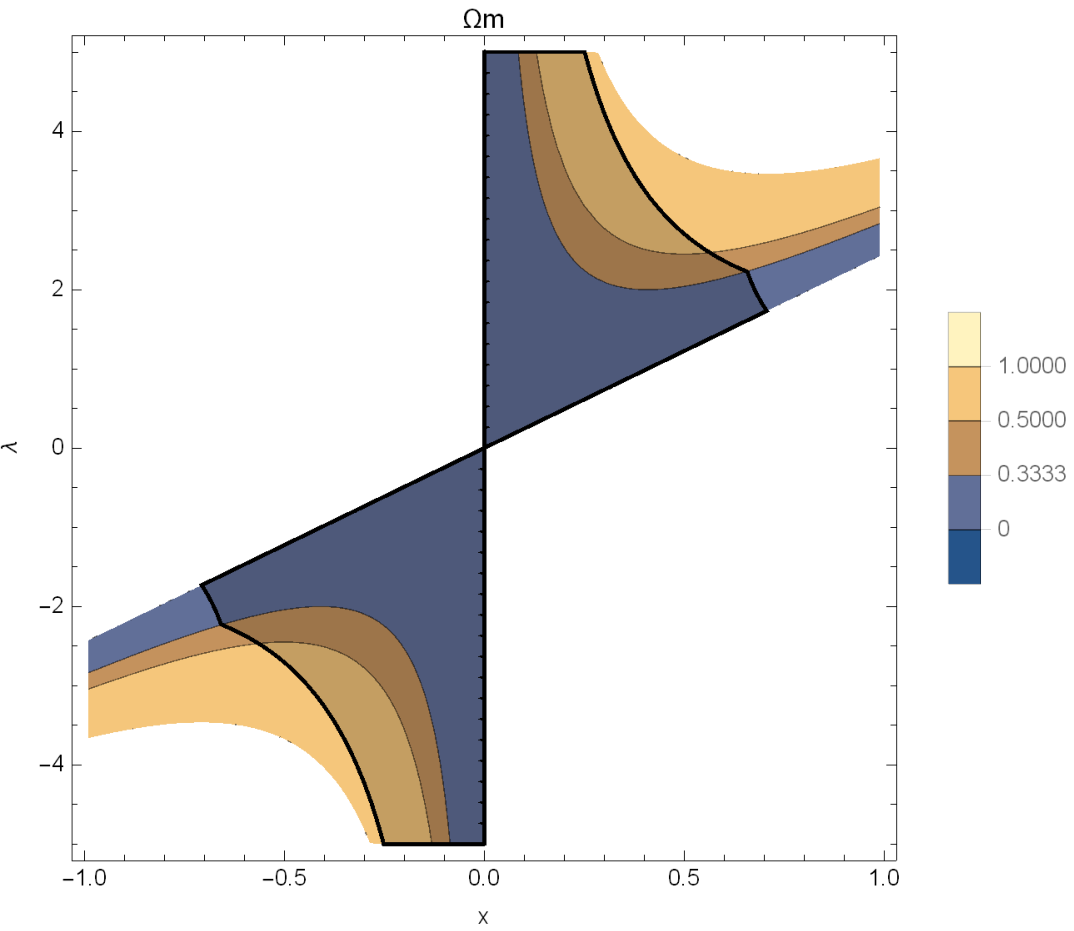}\centering
\caption{Region plot in the space $\left\{ x_{B},\protect\lambda \right\} $
for the interaction constant $Q_{0}$, and the physical parameters $\Omega
_{m},~w_{\protect\phi }$ and $w_{tot}$ for the points which belong to Family
B of the interaction function $\bar{Q}_{B}$. The shaded areas define the
areas where the critical points are stable. }
\label{modelbpb}
\end{figure}

The critical points which correspond to the second family can be one or
three depending on the number of real solutions that the polynomial equation
(\ref{polqb}) admits. We calculate the eigenvalues of the linearized system
and in Fig. \ref{modelbpb} we present the contour diagrams in the space $%
\left\{ x_{B},\lambda \right\} $ for the cosmological parameters $\Omega
_{m}\left( P_{B}\left( \bar{Q}_{B}\right) \right) ,~$ $w_{\phi }\left(
P_{B}\left( \bar{Q}_{B}\right) \right) $ and $w_{tot}\left( P_{B}\left( \bar{%
Q}_{B}\right) \right) $ in which the shaded areas mark the surfaces where
both eigenvalues are negative meaning that the critical points are stable.
Moreover, the contour plot of parameter~$Q_{0}~$is presented.

From Fig. \ref{modelbpb} we can infer that a de Sitter universe can be seen
as a future attractor and the model can describe the late-time acceleration
phase of the universe. However, unstable accelerated eras are not provided
which means that this interacting model cannot explain the early
acceleration phase of the universe.

We now follow a similar fashion as done with earlier interaction model, that
means we aim to investigate the critical points of Family B for a specific
value of the parameter $\lambda $.

\paragraph{Special case $\protect\lambda =2:$}

Consider now that $\lambda =2$. Because we have a fixed value for one of the
parameters we can make plots of $\Omega _{m}\left( P_{B}\left( \bar{Q}%
_{B}\right) \right) ,~$ $w_{\phi }\left( P_{B}\left( \bar{Q}_{B}\right)
\right) $ and $w_{tot}\left( P_{B}\left( \bar{Q}_{B}\right) \right) $ in
terms of the interaction parameter $Q_{0}$. Furthermore, as before, for $%
\lambda =2$, points $P_{B}\left( \bar{Q}_{B}\right) $ exists when $x_{B}\in
(0,\sqrt{2/3}~]$.

\begin{figure}[tbp]
\includegraphics[height=5cm]{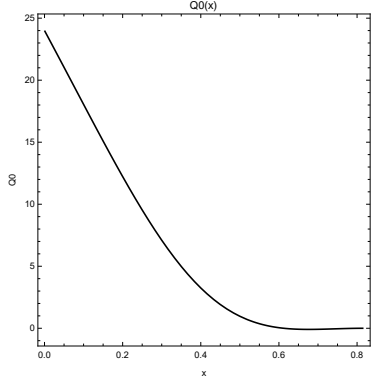}\centering
\caption{Numerical solution for the polynomial equation (\protect\ref{pf.02}%
) }
\label{num0001}
\end{figure}

For $\ \lambda =2$, the algebraic equation (\ref{polqb}) becomes%
\begin{equation}
Q_{0}=6\left( 7-2\sqrt{6}x_{B}-9\left( 3+6x_{B}^{2}-2\sqrt{6}x_{B}\right)
^{-1}\right) ,  \label{pf.02}
\end{equation}%
where from Fig. \ref{num0001} it is clear that for a specific value of $Q_{0}
$, only one point $P_{B}\left( \bar{Q}_{B}\right) $ exists and this becomes
true if and only if $Q_{0}>0.$

\begin{figure}[tbp]
\includegraphics[height=5cm]{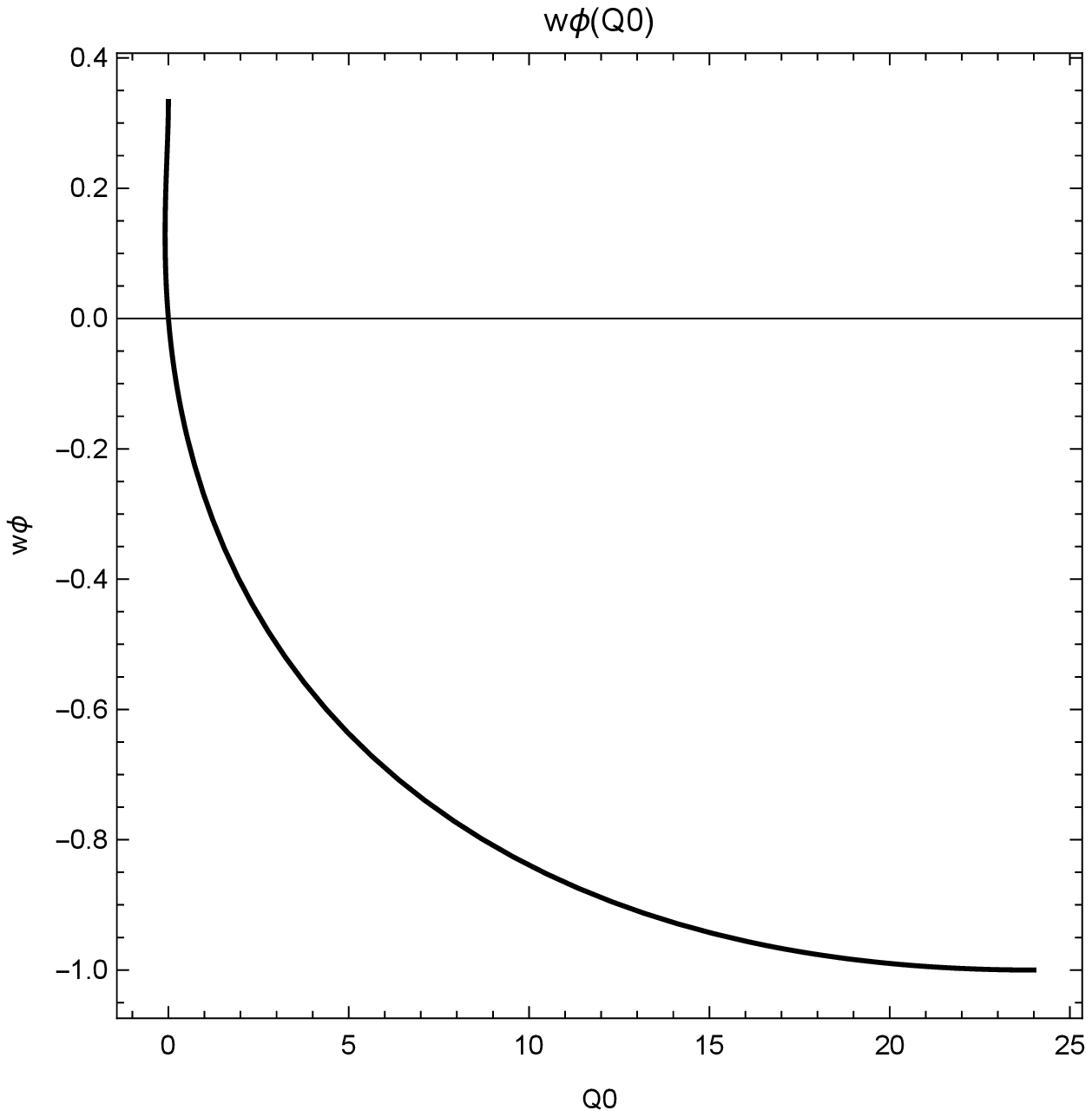}\centering%
\includegraphics[height=5cm]{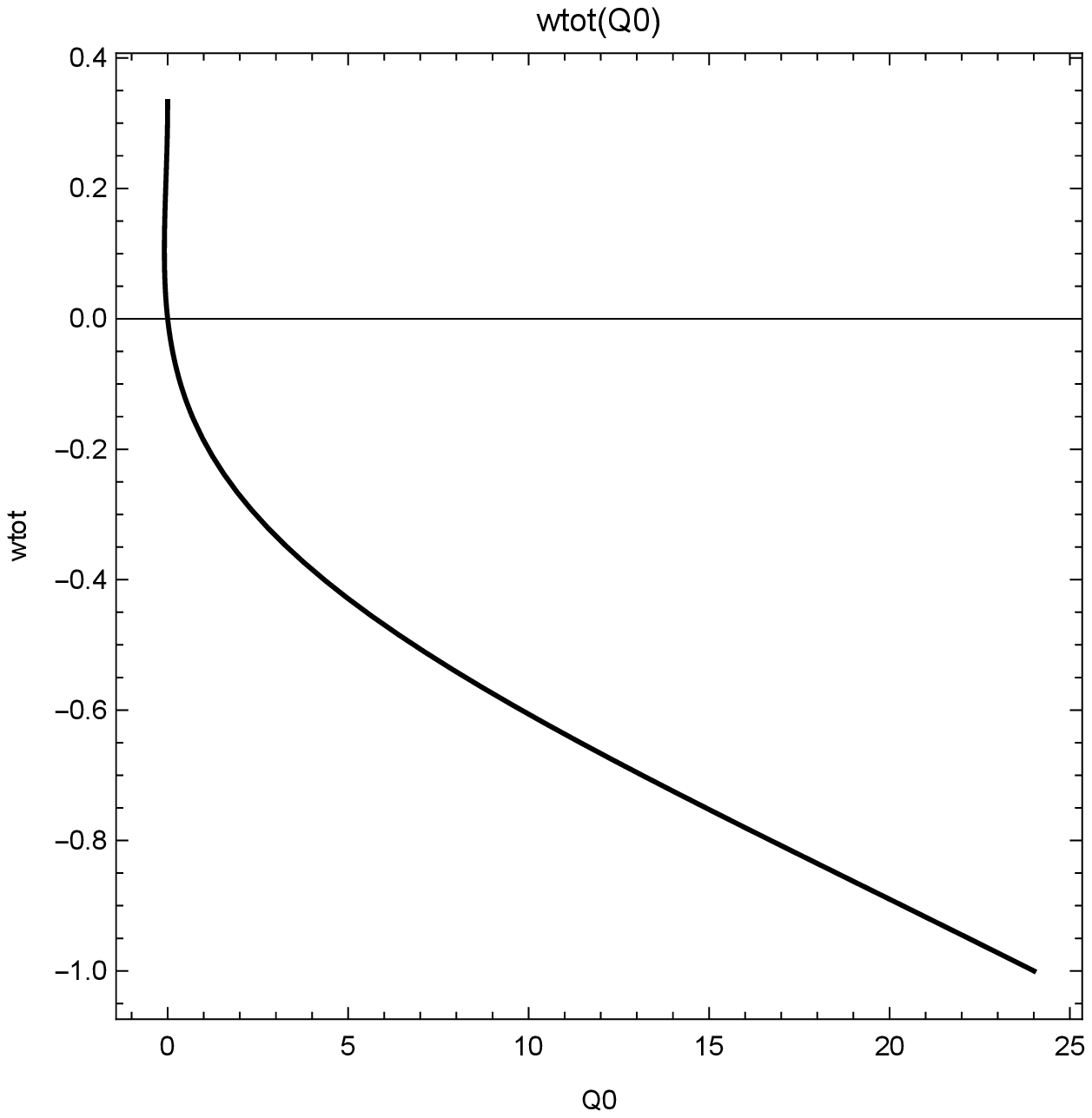}\centering
\newline
\includegraphics[height=5cm]{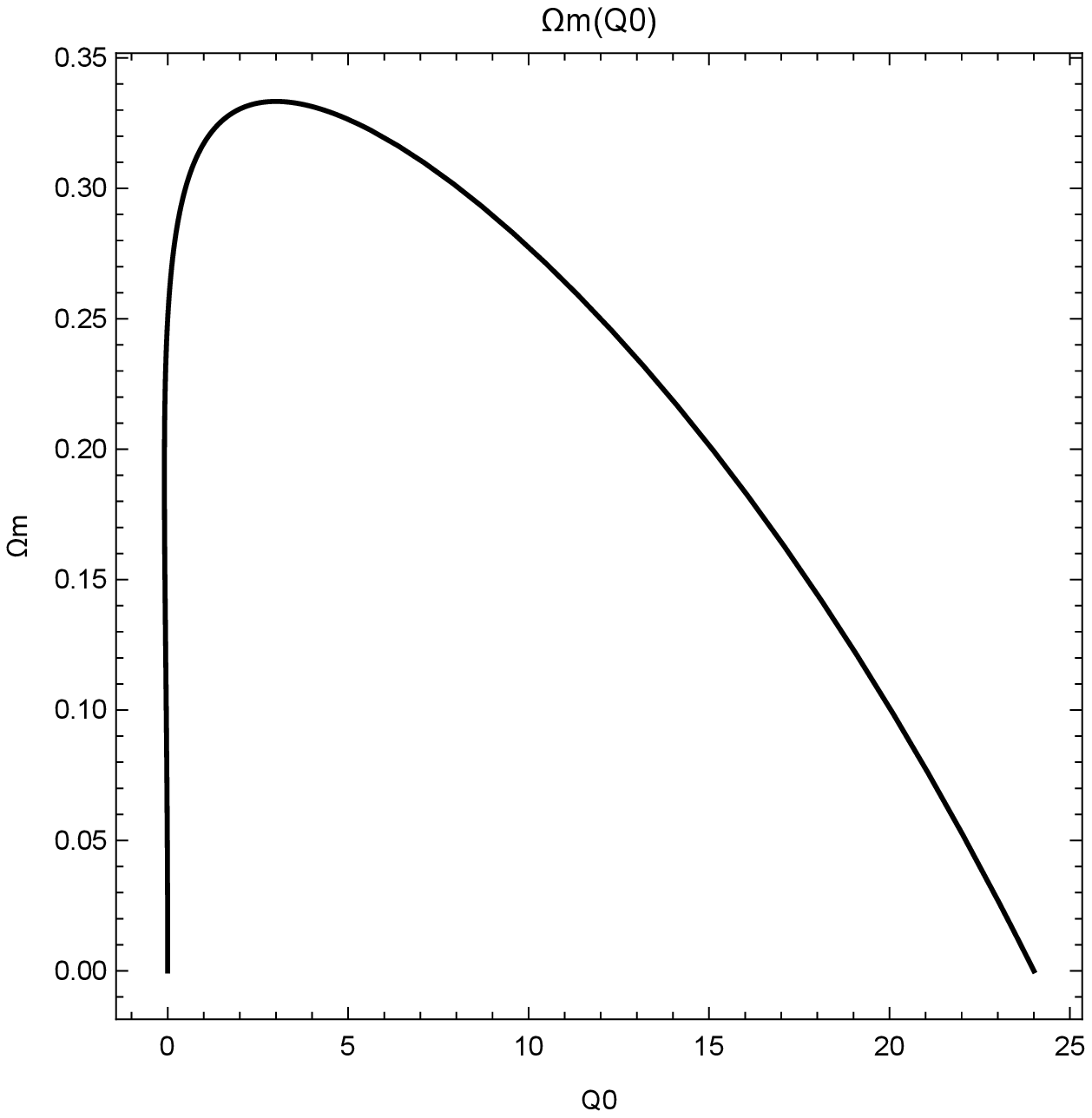}\centering%
\includegraphics[height=5cm]{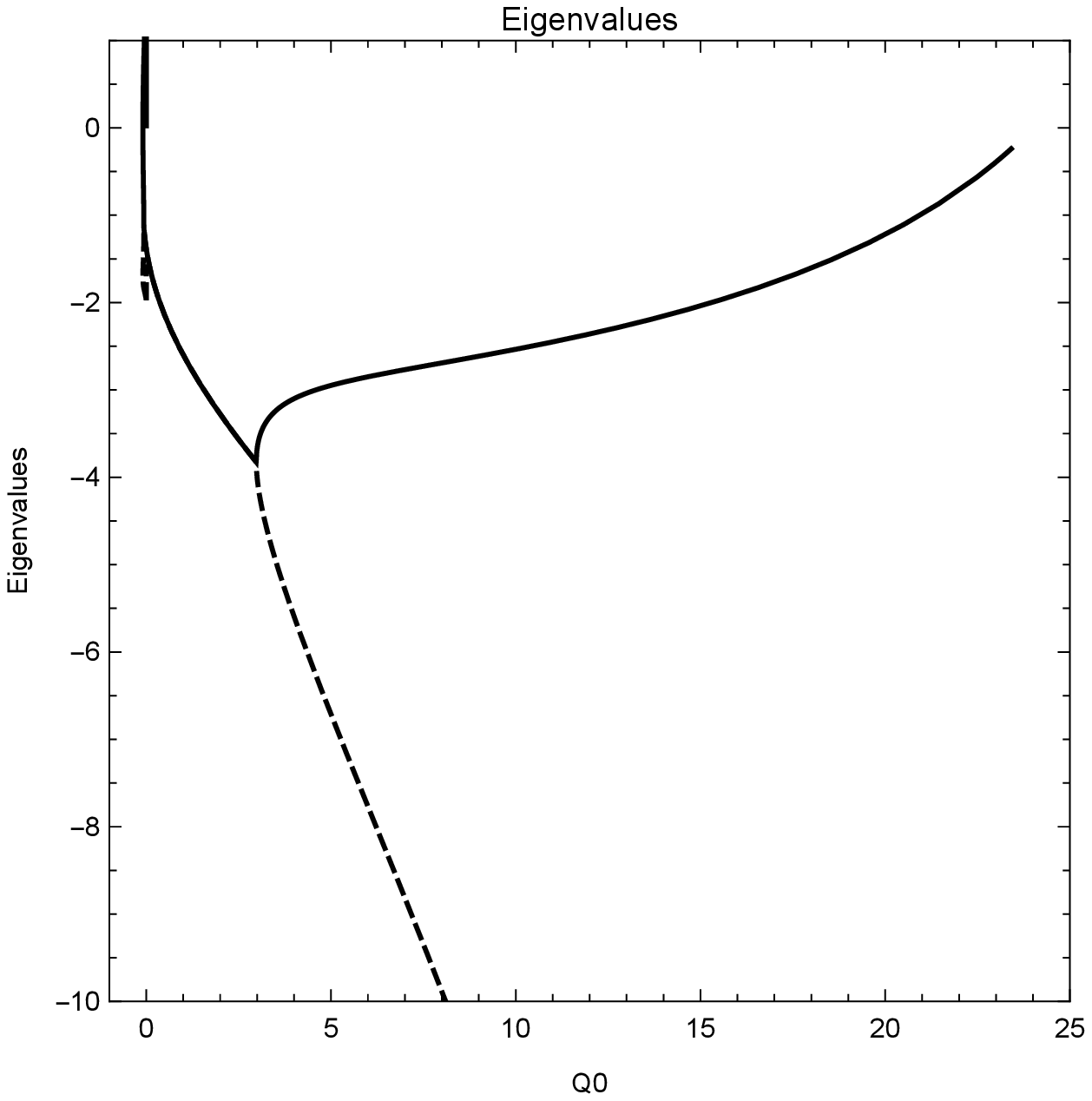}\centering
\caption{Evolution of the physical parameters $w_{\protect\phi }\left(
Q_{0}\right) ,~w_{tot}\left( Q_{0}\right) $ and $\Omega _{m}\left(
Q_{0}\right) $ for the interaction model $Q_{B}$ and $\protect\lambda =2$ at
the critical points of Family B. The evolution of the eigenvalues of the
linearized system is given where we observe that the points are stable for $%
-Q_{0}\gtrsim 0.01$. An unstable radiation-like solution is provided for
values of $Q_{0}<0.01$}
\label{fola20}
\end{figure}

The evolution of the physical parameters in terms of $Q_{0}$ is presented in
Fig. \ref{fola20}. It is clear that for large values of $Q_{0}$ we reach the
de Sitter point, $w_{tot}\left( Q_{0}\right) \rightarrow -1$, while an
inflationary scenario where $\Omega _{m}\neq 0$ and $w_{tot}<-\frac{1}{3}$
is supported by the specific model.

Last but not least, we observe that for small values of $Q_{0},~$i.e. $%
Q_{0}\rightarrow 0$, there exists a radiation-like solution, where the
scalar field can mimic the radiation fluid $w_{\phi }\simeq \frac{1}{3}$,
however, such solution is unstable.

\subsection{Interaction $\bar{Q}_{C}$}

For the interaction $\bar{Q}_{C}$ we find that the dynamical system admits
critical points which belong only to the Family B. More specifically, the
critical points are
\begin{equation}
P_{B}\left( \bar{Q}_{C}\right) =\left( x_{B},\sqrt{1+x_{B}^{2}-\sqrt{\frac{2%
}{3}}\lambda x_{B}}\right)  \label{qcc1}
\end{equation}%
where now $x_{B}$ satisfies the polynomial equation%
\begin{equation}
2x_{B}A\left( x_{B}\right) Q_{0}=3B\left( x_{B}\right) ,  \label{qcc2}
\end{equation}%
in which $A\left( x_{B}\right)$ and $B\left( x_{B}\right)$ are given by
\begin{equation*}
A\left( x_{B}\right) =\left( 6x_{B}^{3}\left( 3x-2\sqrt{6}\lambda \right)
+3\lambda ^{2}-\sqrt{6}\lambda \left( 6+\lambda ^{2}\right) x_{B}+3\left(
6+5\lambda ^{2}\right) x_{B}^{2}\right) ,
\end{equation*}%
\begin{equation*}
B\left( x_{B}\right) =\left(
\begin{array}{c}
4\sqrt{6}\lambda ^{5}x_{B}^{4}-54x_{B}\left( 2x_{B}^{2}+1\right) ^{3}+12%
\sqrt{6}\lambda ^{3}x_{B}^{2}\left( 12x_{B}^{4}+13x_{B}^{2}+3\right) + \\
+9\sqrt{6}\lambda \left( 2x_{B}^{2}+1\right) ^{2}\left(
4x_{B}^{4}+10x_{B}^{2}+1\right) + \\
-48\lambda ^{4}x_{B}^{3}\left( 2x_{B}^{2}+1\right) -36\lambda
^{2}x_{B}\left( 16x_{B}^{6}+30x_{B}^{4}+15x_{B}^{2}+2\right)%
\end{array}%
\right) ,
\end{equation*}%
and it is of order eight in terms of $x_{B}$. That means that the number of
critical points could be 0,~2, 4, 6 or 8, depending on the values of the
free parameters $Q_{0}$ and $\lambda $.

\begin{figure}[tbp]
\includegraphics[height=5cm]{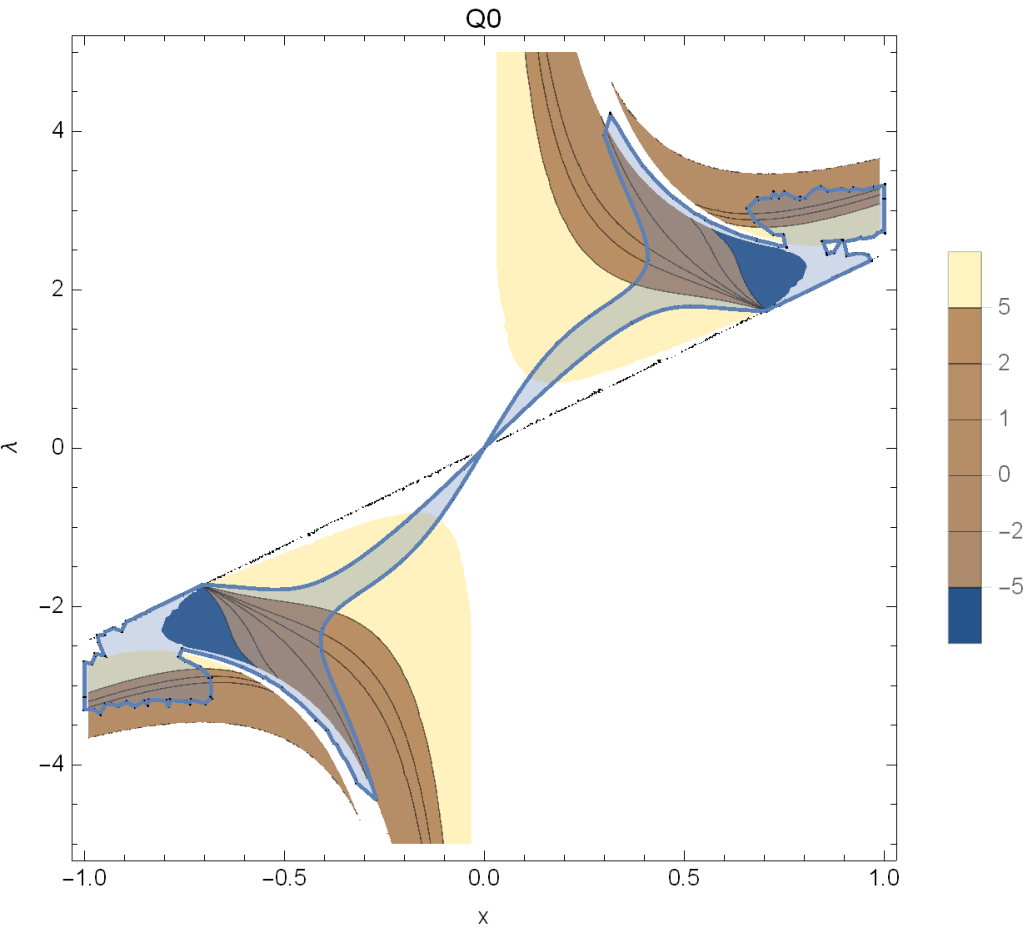}\centering%
\includegraphics[height=5cm]{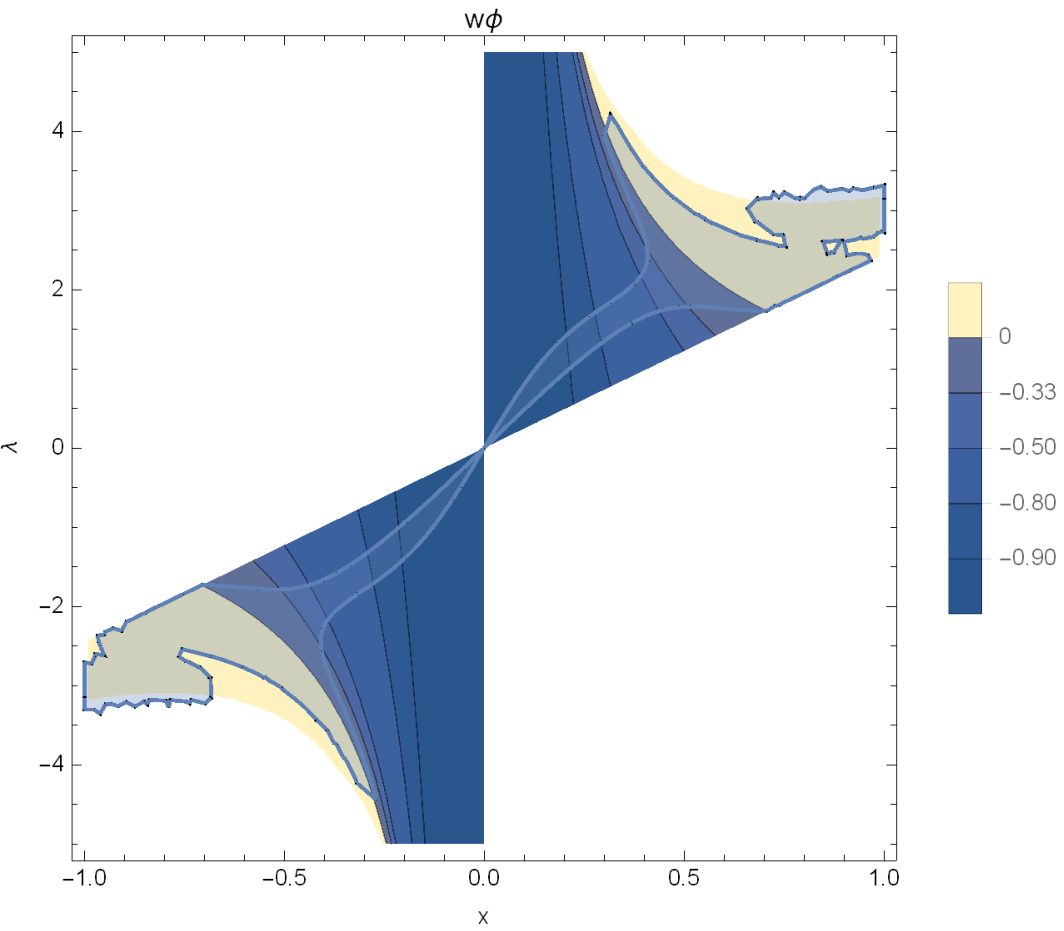}\centering
\newline
\includegraphics[height=5cm]{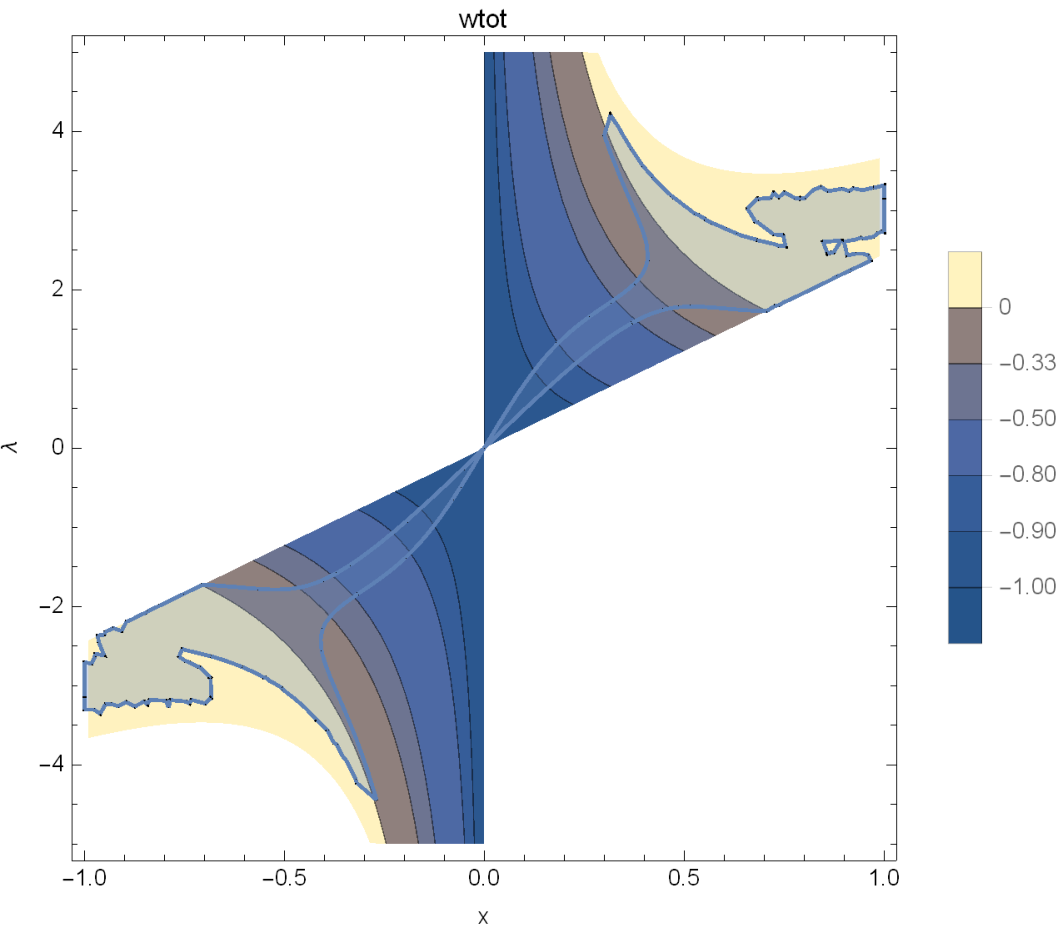}\centering%
\includegraphics[height=5cm]{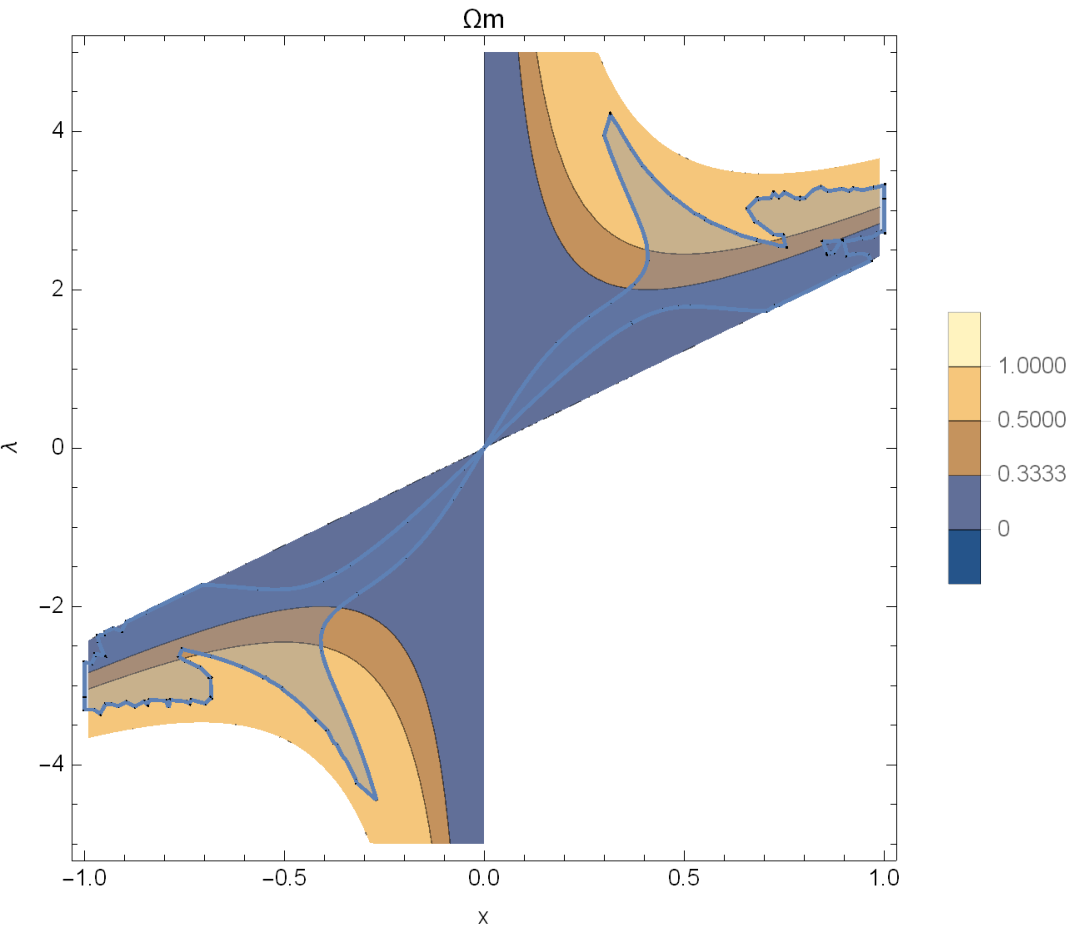}\centering
\caption{Region plot in the space $\left\{ x_{B},\protect\lambda \right\} $
for the interaction constant $Q_{0}$, and the physical parameters $\Omega
_{m},~w_{\protect\phi }$ and $w_{tot}$ for the points which belong to Family
B of the interaction function $\bar{Q}_{C}$. The shaded regions define the
areas where the critical points $P_{B}\left( \bar{Q}_{C}\right) $ are
stable. }
\label{modelpcc1}
\end{figure}

In Fig. \ref{modelpcc1} the numerical solution of the polynomial equation (%
\ref{qcc2}) is presented and the qualitative evolution for different
cosmological parameters namely $w_{tot},~\Omega _{m}$ and $w_{\phi }$ are
also shown. \ The shaded areas in~ Fig. \ref{modelpcc1} indicate that the
real parts of the eigenvalues of the linearized system near to the critical
points are negative, which means that the critical points are stable. It is
straightforward to observe that the present model provides stable
accelerated eras, and also the stable critical points where $\Omega _{m}\neq
0$ and $w_{\phi }<-\frac{1}{3}.$ In order to explain this better we proceed
with the specific case $\lambda =\sqrt{6}.$

\paragraph{Special case $\protect\lambda =\protect\sqrt{6}:$}

We assume that $\lambda =\sqrt{6}$ and now all of our physical parameters
depend only on $Q_{0}$, which means that we can study the one dimensional
parameters $\Omega _{m}\left( P_{B}\left( \bar{Q}_{C}\right) \right) ,~$ $%
w_{\phi }\left( P_{B}\left( \bar{Q}_{C}\right) \right) $ and $w_{tot}\left(
P_{B}\left( \bar{Q}_{C}\right) \right) .$ However, firstly, we need to
determine the number of possible critical points. Indeed from Fig. \ref%
{qcq00l}, we determine that real critical points $P_{B}\left( \bar{Q}%
_{C}\right) $ exist only when $Q_{0}\gtrsim 68$, while the number of
critical points is two.
\begin{figure}[tbp]
\includegraphics[height=5cm]{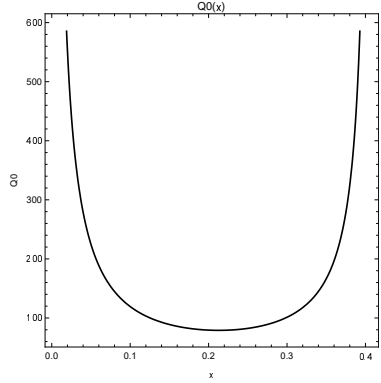}\centering
\caption{Numerical solution for the polynomial equation (\protect\ref{qcc2})
for $\protect\lambda =\protect\sqrt{6}.$ }
\label{qcq00l}
\end{figure}

\begin{figure}[tbp]
\includegraphics[height=5cm]{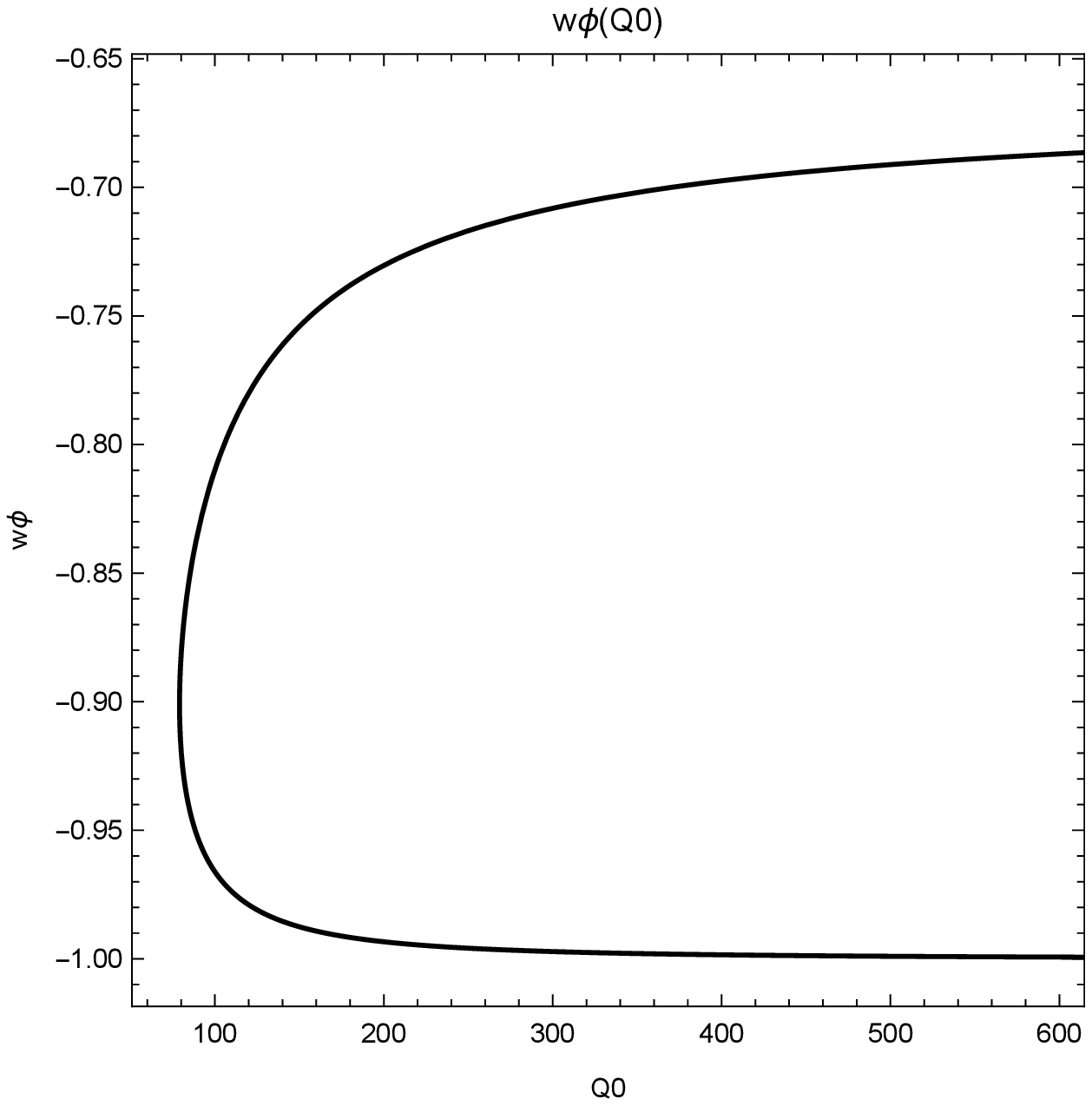}\centering%
\includegraphics[height=5cm]{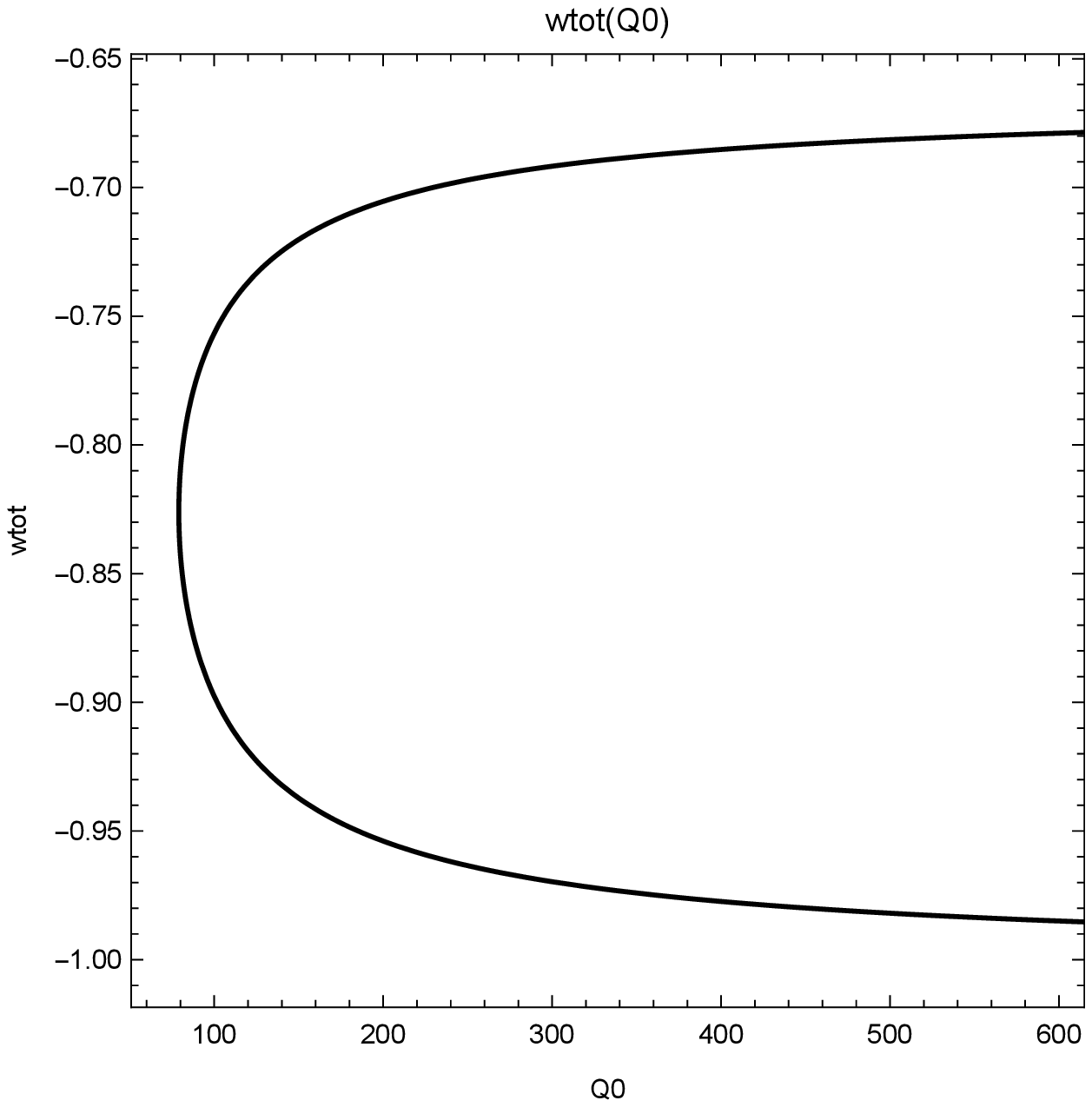}\centering
\newline
\includegraphics[height=5cm]{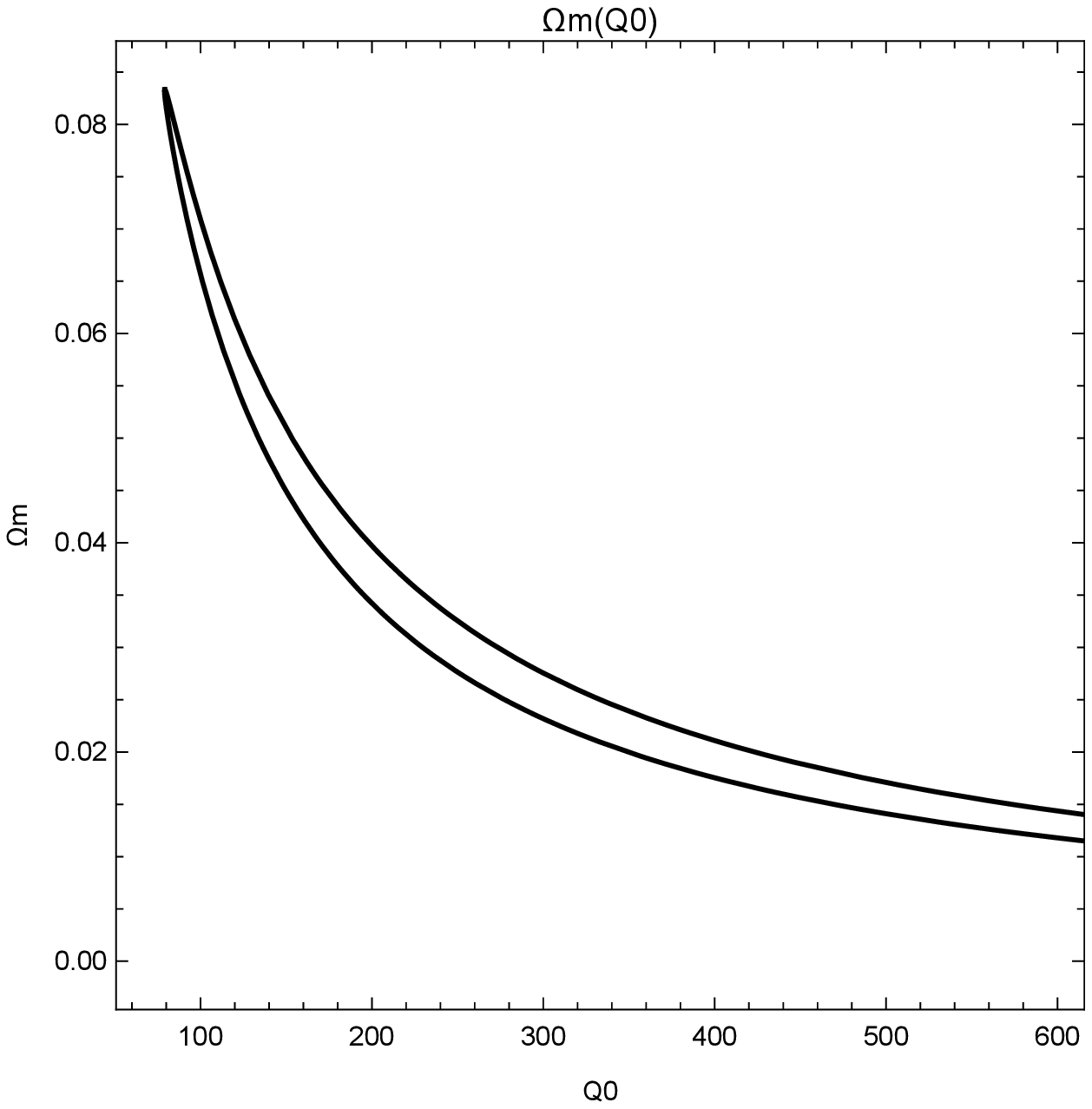}\centering%
\includegraphics[height=5cm]{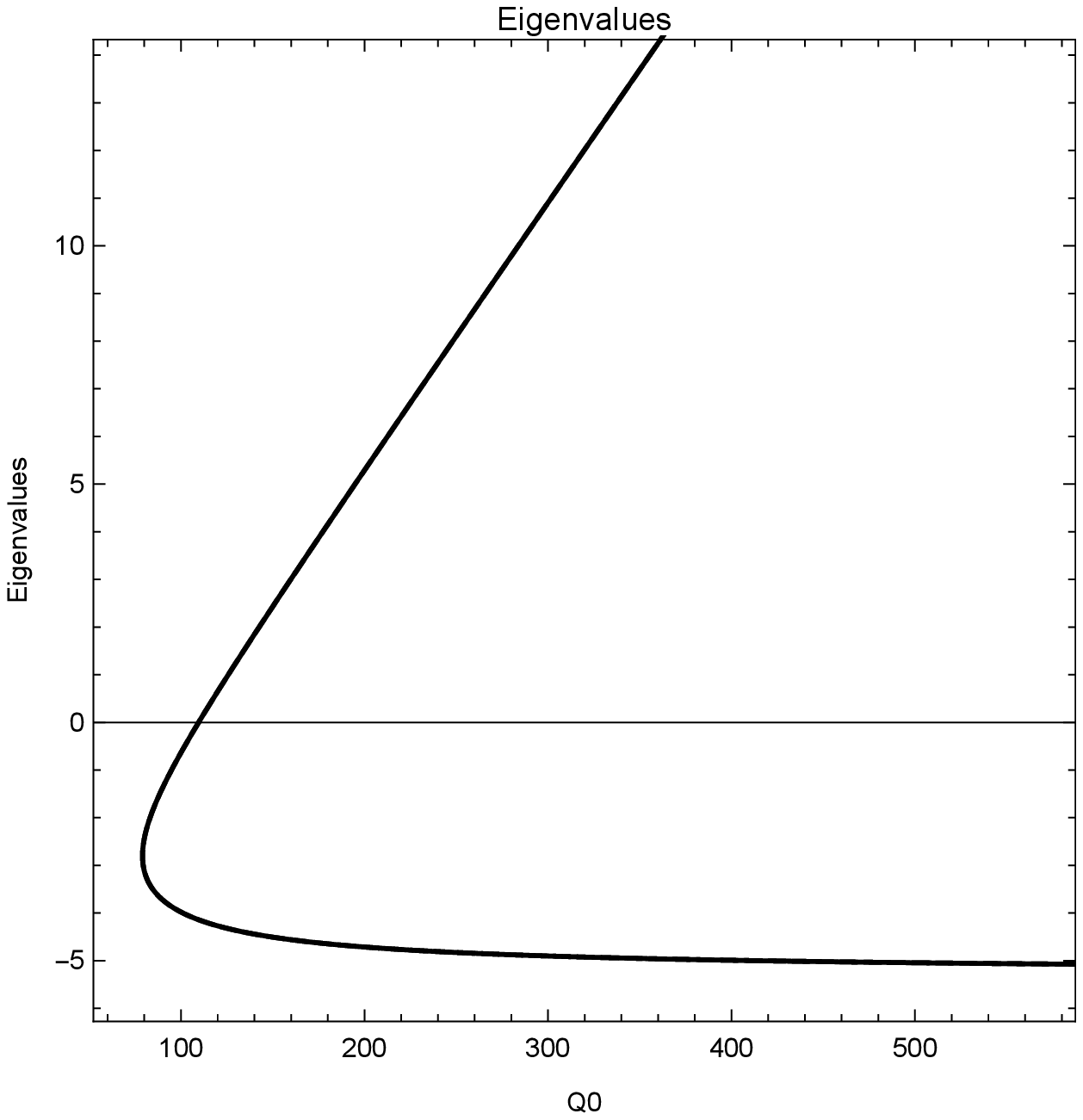}\centering
\caption{Evolution of the physical parameters $w_{\protect\phi }\left(
Q_{0}\right) ,~w_{tot}\left( Q_{0}\right) $ and $\Omega _{m}\left(
Q_{0}\right) $ for the interaction model $Q_{C}$ and $\protect\lambda =%
\protect\sqrt{6}$ at the critical points of Family B. The evolution of the
eigenvalues of the linearized system is given.}
\label{pcc1001}
\end{figure}

As presented in Fig. \ref{pcc1001} the critical points provide that $\Omega
_{m}\left( P_{B}\left( \bar{Q}_{C}\right) \right) \neq 0$ while$~w_{\phi
}\left( P_{B}\left( \bar{Q}_{C}\right) \right) \lesssim -0.7$. In
particular, there are two branches of cosmological solutions with different
cosmological parameters. The one branch is always stable, while the other
branch is stable only for $Q_{0} \gg 300$. Consequently, for smaller values
of that range we can have one stable accelerated solution which approaches
the de Sitter universe and an unstable solution where the scalar field has
an equation of state parameter$~w_{\phi }\left( P_{B}\left( \bar{Q}%
_{C}\right) \right) \lesssim -0.7$ and the dust fluid contributes to the
universe.

The latter observation is important, because that kind of models can provide
$w$CDM(-like) universes. It is important to mention that the limits given in
the Fig. \ref{pcc1001} depend on the value of $\lambda $, hence, other
values of $\lambda $ provide another values for the physical parameters.

\section{Conclusions}

\label{sec-conclu}

Interaction in the dark sector mainly between the dark matter and dark
energy is a possible approach to explain the dynamical features of the
universe. Observational data from various astrophysical sources data have
shown that a mild interaction in the dark sector is allowed. The allowance
of an interaction in the dark sector has been found to explain the tensions
between the cosmological parameters. Additionally, interaction has a well
motivated origin in the cosmological regime. Usually the theory of
interaction rests on the choices of the interaction rates that could be
either linear or nonlinear in nature. The linear interaction rates are
relatively easy compared to the nonlinear models for interaction just
because of their construction and as a consequence this particular kind of
interaction models have got much attention. Since the exact interaction rate
between the dark sectors is not yet known, thus, the cosmological models
allowing nonlinear interaction rates are equally welcome to understand the
dynamics of the universe. The studies on nonlinear interaction rates are
also important to understand its necessity in the cosmological regime.

The present work thus investigates the cosmological dynamics in presence of
various nonlinear interaction rates between dark matter and dark energy in
the background of a spatially flat FLRW universe where the gravitational
sector is described by the Einstein gravity. The dark matter is considered
to be pressureless and dark energy is a minimally coupled scalar field to
gravity and additionally the potential of the scalar field has been assumed
to be exponential: $V(\phi )=V_{0}\exp (-\lambda \phi )$. In particular, we
have considered three distinct nonlinear interaction rates $Q(r,x,y)$ given
in (\ref{models}) where $x$, $y$ are the dimensionless variables defined in (%
\ref{new-variables}) and $r=\frac{\rho _{m}}{\rho _{\phi }}$, is the
coincidence parameter. All interaction models have only one free parameter $%
Q_{0}$, known as interaction parameter or the coupling parameter that
describes the strength of the interaction and the direction of energy
transfer. We then perform the dynamical analysis through the analysis of
critical points and stability. Due to nonlinearity in the models the
dynamics becomes complicated and hence we perform numerical simulation.
Finally we have considered two distinct analysis of the critical points, one
when $y=0$ (Family A) and the one with $y\neq 0$ (Family B) (see section \ref%
{sec-critical}).

For the first nonlinear interaction $\bar{Q}_{A}$, our observations are as
follows. The real critical points of Family A are unstable in nature while
for Family B, we could have stable and unstable real critical points. The
stable critical points cannot describe the present accelerated expansion
(see Fig. \ref{modelapb}) while the de Sitter universe can be obtained as an
unstable solution of this interaction model.

The second nonlinear interaction $\bar{Q}_{B}$ is just the reverse of $\bar{Q%
}_{A}$. Concerning the dynamics of this interaction model, we find that the
critical points belonging to Family A are unstable. Now, for the critical
points of Family B we have some interesting outcomes. For the critical
points of Family B we find that a stable late time accelerated expansion is
possible (see Fig. \ref{modelbpb}) while usually an inflationary solution is
not obtained. However, for specific values of $\lambda $, inflationary
solution is possible. Additionally, for small values of the interaction
parameter $Q_{0}$, radiation-like solution is also admitted.

For the last nonlinear interaction model in this series, namely $\bar{Q}_C$,
we find that the critical points only belong to Family B. Depending on the
free parameters the number of critical points could be one of $\{0, 2, 4, 6,
8\}$. From the numerical simulation (shown in Fig. \ref{modelpcc1}), we find
that the model could allow stable critical points where the accelerated
expansion of the universe is possible. The model has been closely examined
for a specific value of $\lambda = \sqrt{6}$ (see Figs. \ref{qcq00l} and \ref%
{pcc1001}), from which one can see that one stable accelerated expansion
approaching toward the de Sitter universe and one unstable accelerated
expansion are possible. In addition, the model realizes the $w$CDM like
universe. Thus, this interaction model is quite interesting in the
perspective of cosmological dynamics.

Thus, one can see that the nonlinear interaction models are quite appealing
in the context of universe's evolution by offering many possibilities. Since
the interacting dynamics allows us to consider some alternative models,
hence, one may explore the dynamical features with some other models as
well. The existence of stable de Sitter solution is an interesting outcome
of the present nonlinear interaction models.

\begin{acknowledgments}
AP acknowledges financial support of FONDECYT grant no. 3160121. SP
acknowledges the financial support through the Faculty Research and
Professional Development Fund (FRPDF) Scheme of Presidency University,
Kolkata, India. WY thanks the financial support through the National Natural
Science Foundation of China under Grants No. 11705079 and No. 11647153.
\end{acknowledgments}


\begin{thebibliography}{99}
\bibitem{snia1} A.~G.~Riess \textit{et al.} [Supernova Search Team],
Astron.\ J.\ \textbf{116}, 1009 (1998) [astro-ph/9805201].

\bibitem{snia2} S.~Perlmutter \textit{et al.} [Supernova Cosmology Project
Collaboration],
Astrophys.\ J.\ \textbf{517}, 565 (1999) [astro-ph/9812133].

\bibitem{Spergel:2003cb} D.~N.~Spergel \textit{et al.} [WMAP Collaboration],
Astrophys.\ J.\ Suppl.\ \textbf{148}, 175 (2003) 
[astro-ph/0302209].

\bibitem{Tegmark:2006az} M.~Tegmark \textit{et al.} [SDSS Collaboration],
Phys.\ Rev.\ D \textbf{74}, 123507 (2006) 
[astro-ph/0608632].

\bibitem{Spergel:2006hy} D.~N.~Spergel \textit{et al.} [WMAP Collaboration],
Astrophys.\ J.\ Suppl.\ \textbf{170}, 377 (2007) 
[astro-ph/0603449].

\bibitem{Ade:2015xua} P.~A.~R.~Ade \textit{et al.} [Planck Collaboration],
Astron.\ Astrophys.\ \textbf{594}, A13 (2016)
[arXiv:1502.01589 [astro-ph.CO]].

\bibitem{Amendola:2016saw} L.~Amendola \textit{et al.},
Living Rev.\ Rel.\ \textbf{21}, no. 1, 2 (2018)
[arXiv:1606.00180 [astro-ph.CO]].

\bibitem{Aghanim:2018eyx} N.~Aghanim \textit{et al.} [Planck Collaboration],
[arXiv:1807.06209 [astro-ph.CO]]. 

\bibitem{Abbott:2018wog} T.~M.~C.~Abbott \textit{et al.} [DES
Collaboration],
arXiv:1811.02374 [astro-ph.CO].

\bibitem{Amendola-ide1} L. Amendola, 
Phys. Rev. D \textbf{62}, 043511 (2000) [arXiv:astro-ph/9908023].

\bibitem{Amendola-ide2} L. Amendola and C. Quercellini,
Phys. Rev. D \textbf{68}, 023514 (2003) [arXiv:astro-ph/0303228].

\bibitem{Pavon:2005yx} D.~Pav\'{o}n and W.~Zimdahl,
Phys.\ Lett.\ B \textbf{628}, 206 (2005) [arXiv:gr-qc/0505020].

\bibitem{delCampo:2008sr} S.~del Campo, R.~Herrera and D.~Pav\'{o}n,
Phys.\ Rev.\ D \textbf{78}, 021302 (2008) [arXiv:0806.2116 [astro-ph]].

\bibitem{delCampo:2008jx} S.~del Campo, R.~Herrera and D.~Pav\'{o}n,
J. Cosmol. Astropart. Phys. \textbf{0901}, 020 (2009) [arXiv:0812.2210
[gr-qc]].

\bibitem{Wetterich-ide1} C. Wetterich,
Astron. Astrophys. \textbf{301}, 321 (1995) [arXiv:hep-th/9408025].

\bibitem{Yang:2014gza} W.~Yang and L.~Xu,
Phys.\ Rev.\ D \textbf{89}, no. 8, 083517 (2014)
[arXiv:1401.1286 [astro-ph.CO]].

\bibitem{Salvatelli:2014zta} V.~Salvatelli, N.~Said, M.~Bruni, A.~Melchiorri
and D.~Wands,
Phys.\ Rev.\ Lett.\ \textbf{113}, no. 18, 181301 (2014)
[arXiv:1406.7297 [astro-ph.CO]].

\bibitem{Yang:2014hea} W.~Yang and L.~Xu,
Phys.\ Rev.\ D \textbf{90}, no. 8, 083532 (2014)
[arXiv:1409.5533 [astro-ph.CO]].

\bibitem{Nunes:2016dlj} R.~C.~Nunes, S.~Pan and E.~N.~Saridakis,
Phys.\ Rev.\ D \textbf{94}, 023508 (2016) [arXiv:1605.01712 [astro-ph.CO]].

\bibitem{Kumar:2016zpg} S.~Kumar and R.~C.~Nunes,
Phys. Rev. D \textbf{94}, 123511 (2016) [arXiv:1608.02454 [astro-ph.CO]].

\bibitem{Yang:2016evp} W.~Yang, H.~Li, Y.~Wu and J.~Lu,
JCAP \textbf{1610}, no.10, 007 (2016) 
[arXiv:1608.07039 [astro-ph.CO]].

\bibitem{Kumar:2017dnp} S.~Kumar and R.~C.~Nunes,
Phys.\ Rev.\ D \textbf{96}, no. 10, 103511 (2017)
[arXiv:1702.02143 [astro-ph.CO]].

\bibitem{DiValentino:2017iww} E.~Di Valentino, A.~Melchiorri and O.~Mena,
arXiv:1704.08342 [astro-ph.CO].

\bibitem{Yang:2017yme} W.~Yang, N.~Banerjee and S.~Pan,
Phys.\ Rev.\ D \textbf{95}, 123527 (2017) 
[arXiv:1705.09278 [astro-ph.CO]].

\bibitem{Billyard:2000bh} A.~P.~Billyard and A.~A.~Coley,
Phys.\ Rev.\ D \textbf{61}, 083503 (2000), 
arXiv:astro-ph/9908224

\bibitem{Barrow:2006hia} J.~D.~Barrow and T.~Clifton,
Phys.\ Rev.\ D \textbf{73}, 103520 (2006), 
[gr-qc/0604063]

\bibitem{Amendola:2006dg} L.~Amendola, G.~Camargo Campos and R.~Rosenfeld,
Phys.\ Rev.\ D \textbf{75}, 083506 (2007), 
[astro-ph/0610806]

\bibitem{CalderaCabral:2008bx} G. Caldera-Cabral, R. Maartens, L.A. Ure\~{n}%
a-L\'{o}pez, 
Phys.\ Rev.\ D \textbf{79}, 063518 (2009), 
[arXiv:0812.1827 [gr-qc]]

\bibitem{Quartin:2008px} M.~Quartin, M.~O.~Calvao, S.~E.~Joras,
R.~R.~R.~Reis and I.~Waga,
JCAP \textbf{0805}, 007 (2008), 
arXiv:0802.0546 [astro-ph]

\bibitem{Valiviita:2009nu} J.~V\"{a}liviita, R.~Maartens and E.~Majerotto,
Mon.\ Not.\ Roy.\ Astron.\ Soc.\ \textbf{402}, 2355 (2010),
arXiv:0907.4987 [astro-ph.CO]

\bibitem{Clemson:2011an} T.~Clemson, K.~Koyama, G.~B.~Zhao, R.~Maartens and
J.~Valiviita,
Phys.\ Rev.\ D \textbf{85}, 043007 (2012), 
[arXiv:1109.6234 [astro-ph.CO]]

\bibitem{Thorsrud:2012mu} M.~Thorsrud, D.~F.~Mota and S.~Hervik,
JHEP \textbf{1210}, 066 (2012), 
[arXiv:1205.6261 [hep-th]]

\bibitem{Pan:2012ki} S.~Pan, S.~Bhattacharya and S.~Chakraborty,
Mon.\ Not.\ Roy.\ Astron.\ Soc.\ \textbf{452}, 3038 (2015),
arXiv:1210.0396 [gr-qc]

\bibitem{Pan:2016ngu} S.~Pan and G.~S.~Sharov,
Mon.\ Not.\ Roy.\ Astron.\ Soc.\ \textbf{472}, 4736 (2017),
[arXiv:1609.02287 [gr-qc]]

\bibitem{Sharov:2017iue} G.~S.~Sharov, S.~Bhattacharya, S.~Pan, R.~C.~Nunes
and S.~Chakraborty,
Mon.\ Not.\ Roy.\ Astron.\ Soc.\ \textbf{466}, no. 3, 3497 (2017),
[arXiv:1701.00780 [gr-qc]]

\bibitem{Pan:2017ent} S.~Pan, A.~Mukherjee and N.~Banerjee,
Mon.\ Not.\ Roy.\ Astron.\ Soc.\ \textbf{477}, 1189 (2018),
[arXiv:1710.03725 [astro-ph.CO]]

\bibitem{Yang:2018euj} W.~Yang, S.~Pan, E.~Di Valentino, R.~C.~Nunes,
S.~Vagnozzi and D.~F.~Mota,
JCAP \textbf{1809}, no. 09, 019 (2018) 
[arXiv:1805.08252 [astro-ph.CO]].

\bibitem{Yang:2018pej} W.~Yang, S.~Pan and A.~Paliathanasis,
Mon.\ Not.\ Roy.\ Astron.\ Soc.\ \textbf{482}, 1007 (2019)
[arXiv:1804.08558 [gr-qc]].

\bibitem{Yang:2018ubt} W.~Yang, S.~Pan, L.~Xu and D.~F.~Mota,
Mon.\ Not.\ Roy.\ Astron.\ Soc.\ \textbf{482}, 1858 (2019)
[arXiv:1804.08455 [astro-ph.CO]].

\bibitem{int1} B. Wang, E. Abdalla, F. Atrio-Barandela and D. Pavon, Pept.
Prog. Phys \textbf{79,} 096901 (2016)

\bibitem{int2} E. Majerotto, J. Valiviita and R. Marrtens, Nucl. Phys. B
\textbf{194,} 260 (2006)

\bibitem{int3} H. Zonunmavia, W. Khyllep, N. Roy, J. Dutta and N.\ Tamanini,
Phys.\ Rev. D \textbf{96,} 083527 (2017)

\bibitem{int4} G. Papagiannopoulos, S. Basilakos, A. Paliathanasis, S.
Savvidou and P.C.\ Stavrinos, Class. Quant. Grav.\textbf{\ 34,} 225008 (2017)

\bibitem{int5} G. Leon, A. Paliathanasis and J.L. Morales-Martinez, Eur.
Phys. J. C \textbf{78}, 753 (2018)

\bibitem{int6} S.Kr. Biswas, W. Khyllep, J. Dutta and S. Chakraborty, Phys.\
Rev. D \textbf{95}, 103009 (2017)

\bibitem{int7} P.Tsiapi and S.\ Basilakos, Mon.\ Not.\ Roy.\ Astron.\ Soc.
10.1093/mnras/stz540 [arXiv:1810.12902]

\bibitem{Pavon:2007gt} D.~Pav\'{o}n and B.~Wang,
Gen.\ Rel.\ Grav.\ \textbf{41}, 1 (2009) 
[arXiv:0712.0565 [gr-qc]].

\bibitem{Chimento:2009hj} L.~P.~Chimento,
Phys.\ Rev.\ D \textbf{81}, 043525 (2010) 
[arXiv:0911.5687 [astro-ph.CO]].

\bibitem{Arevalo:2011hh} F.~Arevalo, A.~P.~R.~Bacalhau and W.~Zimdahl,
Class.\ Quant.\ Grav.\ \textbf{29}, 235001 (2012)
[arXiv:1112.5095 [astro-ph.CO]].

\bibitem{Yang:2017zjs} W.~Yang, S.~Pan and J.~D.~Barrow,
Phys.\ Rev.\ D \textbf{97}, no. 4, 043529 (2018)
[arXiv:1706.04953 [astro-ph.CO]].

\bibitem{Copeland:2006wr} E.~J.~Copeland, M.~Sami and S.~Tsujikawa,
Int.\ J.\ Mod.\ Phys.\ D \textbf{15}, 1753 (2006),
[hep-th/0603057].

\bibitem{Copeland:1997et} E.~J.~Copeland, A.~R.~Liddle and D.~Wands,
Phys.\ Rev.\ D \textbf{57}, 4686 (1998) 
[gr-qc/9711068].
\end{thebibliography}
\end{document}